\documentclass[12pt]{article}
\usepackage{epsf}
\usepackage{amsmath,amssymb}

\usepackage[dvips]{graphicx}

\usepackage{comment}

\begin{document}

\newcommand{\lsim}{\stackrel{<}{_\sim}}
\newcommand{\gsim}{\stackrel{>}{_\sim}}
\newcommand{\mathhyphen}{\mathchar"712D}

\renewcommand{\theequation}{\thesection.\arabic{equation}}

\renewcommand{\thefootnote}{\fnsymbol{footnote}}
\setcounter{footnote}{0}

\begin{titlepage}

\begin{center}

\hfill KEK-TH-1971\\
\hfill UT-17-10\\
\hfill April, 2017\\

\vskip .75in

{\Large\bf
  False Vacuum Decay in Gauge Theory
}

\vskip .5in

{\large
  Motoi Endo$^{\rm (a,b,c)}$, Takeo Moroi$^{\rm (d,c)}$, 
  Mihoko M. Nojiri$^{\rm (a,b,c)}$, Yutaro Shoji$^{\rm (e)}$
}

\vskip 0.25in

$^{\rm (a)}${\em 
KEK Theory Center, IPNS, KEK, Tsukuba, Ibaraki 305-0801, Japan}

\vskip 0.1in
$^{\rm (b)}${\em 
The Graduate University of Advanced Studies (Sokendai),\\
Tsukuba, Ibaraki 305-0801, Japan}

\vskip 0.1in
$^{\rm (c)}${\em 
Kavli IPMU (WPI), University of Tokyo, Kashiwa, Chiba 277-8583, Japan}

\vskip 0.1in
$^{\rm (d)}${\em 
Department of Physics, University of Tokyo, Tokyo 113-0033, Japan}

\vskip 0.1in
$^{\rm (e)}${\em 
Institute for Cosmic Ray Research, The University of Tokyo, 
Kashiwa 277-8582, Japan}

\end{center}

\vskip .5in

\begin{abstract}
  
  The decay rate of a false vacuum is studied in gauge theory, paying
  particular attention to its gauge invariance.  Although the decay
  rate should not depend on the gauge parameter $\xi$ according to the
  Nielsen identity, the gauge invariance of the result of a perturbative
  calculation has not been clearly shown.  We give a prescription to
  perform a one-loop calculation of the decay rate, with which a
  manifestly gauge-invariant expression of the decay rate is obtained.
  We also discuss the renormalization necessary to make the result finite,
  and show that the decay rate is independent of the gauge parameter
  even after the renormalization.
  
\end{abstract}

\end{titlepage}

\setcounter{page}{1}
\renewcommand{\thefootnote}{\#\arabic{footnote}}
\setcounter{footnote}{0}

\section{Introduction}
\label{sec:intro}
\setcounter{equation}{0}

Calculation of the decay rate of a false vacuum (i.e., bubble
nucleation rate) was formulated in \cite{Coleman:1977py,
  Callan:1977pt, Coleman:aspectsof} by introducing the so-called   bounce,
a saddle-point solution of the Euclidean classical equation of motion. The
decay rate of the false vacuum per unit volume is expressed as
\begin{align}
  \gamma = {\cal A} e^{-{\cal B}}.
  \label{decayrate}
\end{align}
Here, ${\cal B}$ is the bounce action, the Euclidean classical action
of the bounce configuration.  The prefactor ${\cal A}$ is obtained by
integrating out field fluctuations around the bounce configuration as
well as those around the false vacuum.  It takes account of radiative
corrections (i.e., loop corrections) to the effective action of the
bounce.

When a scalar field responsible for the metastability of the false
vacuum has gauge interactions, fluctuations of the gauge fields as well
as those of the Faddeev-Popov (FP) ghosts contribute to the prefactor
${\cal A}$.  Gauge dependence of ${\cal A}$ is the main subject of
this paper.  Gauge fixing is necessary for the calculation of the
radiative corrections, with which a gauge parameter (which will be
called $\xi$ in our analysis) is introduced.  Then, some of the
propagators of the fields acquire unphysical poles which depend on
$\xi$; the $\xi$-dependence should vanish from physical quantities.
According to the Nielsen identity, the effective action is gauge
independent at its extrema \cite{Nielsen:1975fs, Fukuda:1975di},
although, in general, the effective action is gauge dependent.  In
perturbative calculations of the decay rate of the false vacuum,
however, it has not been clarified how the gauge dependence vanishes
and what the gauge invariant expression of the decay rate is.

The prefactor ${\cal A}$ consists of functional determinants of
second-order differential operators (so-called fluctuation operators)
governing mode functions of the field fluctuations.  Such functional
determinants are expressed by asymptotic values of solutions of the
second-order differential equations.  Evaluation of the functional
determinants has several complications in gauge theories.  First,
gauge and Nambu-Goldstone (NG) fields mix with each other when the
gauge symmetry is spontaneously broken, which makes the behavior of the
solutions complicated.  Second, as we have already mentioned, the
fluctuation operators contain the gauge parameter $\xi$ so that the
$\xi$-independence of the decay rate is not manifest.  These make it
difficult not only to understand the gauge invariance but also to
numerically calculate the decay rate.  Indeed, for a stable numerical
calculation, the 't~Hooft-Feynman gauge with $\xi=1$ is usually
adopted, with which the fluctuation operators become simple.  However,
with a calculation based on a particular choice of the gauge
parameter, the gauge invariance of the result can not be discussed
directly.

Recently, a gauge-invariant expression of the decay rate has been
derived for a case where gauge symmetry is spontaneously broken in the
false vacuum \cite{Endo:2017gal}.  In \cite{Endo:2017gal}, a gauge
fixing function which reduces to the $R_\xi$ gauge around the false
vacuum has been adopted.  (We call such a gauge as an $R_\xi$-like
gauge.)  However the procedure proposed in \cite{Endo:2017gal} cannot
be applied if gauge symmetry is preserved in the false vacuum.  This
is because, in such a case, there shows up a class of bounce
configurations related by the internal symmetry, all of which
contribute to the false vacuum decay.  With the $R_\xi$-like gauge
fixing, the fluctuation operators are dependent on the bounce
configuration, which makes it difficult to take account of effects of
all the possible bounce configurations.

In this paper, we study the decay of the false vacuum in 4-dimensional
(4D) gauge theory, paying particular attention to the gauge invariance
of the decay rate.\footnote
{For the study of the thermal transition rate of sphaleron, see
  \cite{Baacke:1999sc}.}
We improve the analysis of \cite{Endo:2017gal} and present a
prescription giving rise to a gauge-invariant expression of the decay
rate, which is applicable to the symmetry-preserving false vacuum.  We
use the following gauge fixing function, ${\cal F}=\partial_\mu A_\mu$
(with $A_\mu$ being the gauge field), with which we show that the
difficulty mentioned above can be avoided.  We resolve the
complication due to the mixing between the gauge and NG modes, and
systematically integrate out the fluctuations of the gauge field, NG
mode, and the FP ghosts to calculate the prefactor ${\cal A}$.  We
give a manifestly gauge invariant expression of the decay rate of the
false vacuum.  Our results are useful not only for understanding the
gauge invariance of the decay rate but also for simplifying the numerical
calculation of the decay rate.

The organization of this paper is as follows.  In Section
\ref{sec:setup}, we describe the model we consider.  Important
formulae for our analysis are discussed in Section \ref{sec:formulae}.
Calculations of the functional determinants for the cases with and
without the spontaneous symmetry breaking of the gauge symmetry at the
false vacuum are given in Sections \ref{sec:nonzerovev} and
\ref{sec:zerovev}, respectively.  The issues related to the
renormalization are studied in \ref{sec:renormalization}.  The final
expression of the decay rate can be derived from Eqs.\
\eqref{Atot(final)} $-$ \eqref{Ac(final)}; readers who are interested
only in the result can skip to these equations.  Section
\ref{sec:conclusion} is devoted for conclusions and discussion.

\section{Set Up}
\label{sec:setup}
\setcounter{equation}{0}

\subsection{Lagrangian, bounce}

We consider a model with $U(1)$ gauge symmetry; application of our
argument to the case with non-abelian gauge groups is straightforward.
For simplicity, we concentrate on the case where there exists only one
charged scalar field $\Phi$ which acquires a vacuum expectation value
(VEV).  The Euclidean Lagrangian is given in the following form:
\begin{align}
  {\cal L} = \frac{1}{4} F_{\mu\nu} F_{\mu\nu} 
  + [ (\partial_\mu + i g A_\mu) \Phi^\dagger ]
  [ (\partial_\mu - i g A_\mu) \Phi ]
  + V + {\cal L}_{\rm G.F.} + {\cal L}_{\rm ghost},
  \label{Ltot}
\end{align}
where $F_{\mu\nu}=\partial_\mu A_\nu-\partial_\nu A_\mu$, and $V$ is
the scalar potential.  In addition, ${\cal L}_{\rm G.F.}$ and ${\cal
  L}_{\rm ghost}$ are the gauge fixing terms and the Lagrangian of the
FP ghosts, respectively.  The scalar potential $V$ has true and false
vacua.  We denote the scalar amplitude at the false vacuum to be
$v/\sqrt{2}$, and choose the field configuration of the false vacuum
as
\begin{align}
  (A_\mu,\Phi)_{\rm false\, vacuum} = (0,v/\sqrt{2}).
\end{align}

In the path integral formulation, the decay of the false vacuum is
dominated by a classical path, i.e., an $O(4)$ symmetric solution of
the 4D Euclidean classical equation of motion with appropriate
boundary conditions (the so-called  bounce) \cite{Coleman:1977py,
  Callan:1977pt, Coleman:aspectsof}.  The bounce solution settles to
the false vacuum at the infinity of the Euclidean space:
\begin{align}
  \lim_{r\rightarrow\infty} (A_\mu,\Phi)_{\rm bounce} = (0,v/\sqrt{2}),
  \label{bc_bounce}
\end{align}
where $r\equiv\sqrt{x_\mu x_\mu}$ is the 4D radius in the Euclidean
space.  The bounce is characterized by the function $\bar{\phi}(r)$
which obeys
\begin{align}
  \left[
    \partial_r^2 \Phi + \frac{3}{r} \partial_r \Phi
    - V_\Phi
  \right]_{\Phi\rightarrow\bar{\phi}/\sqrt{2}}
  = 0,
  \label{ClassicalEq}
\end{align}
where $V_\Phi\equiv\partial V/\partial\Phi$.  It also satisfies
\begin{align}
  &\partial_r \bar{\phi} (r=0) = 0,
  \\
  &\bar{\phi} (r=\infty) = v,
\end{align}
where the center of the bounce is set to be $r=0$.  We assume that
$\bar{\phi}$ is a real function of $r$.  The bounce configuration is
given by $(A_\mu,\Phi)=(0,\bar{\phi}/\sqrt{2})$ when $v\neq 0$, while
that for the case of $v=0$ is not unique as we will explain later.

It is important to understand asymptotic behavior of the function
$\bar{\phi}$.  Let us assume that the leading term of the scalar
potential around the false vacuum is quadratic.  Then, because
$\bar{\phi}$ settles to the false-vacuum amplitude at
$r\rightarrow\infty$, the asymptotic behavior of $\bar{\phi}$ can be
understood by using the following equation:
\begin{align}
  \partial_r^2 \bar{\phi} + \frac{3}{r} \partial_r \bar{\phi} -
  m_\phi^2 (\bar{\phi}-v) \simeq 0,
\end{align}
where $m_\phi^2$ is the curvature of the potential around the false
vacuum.  We parameterize the asymptotic behavior of $\bar{\phi}$ as
\begin{align}
  \bar{\phi} (r\rightarrow\infty) 
  \simeq v + \kappa \frac{e^{-m_\phi r}}{r^{3/2}},
\end{align}
where $\kappa$ is a constant.

In some of the previous studies \cite{Isidori:2001bm, Alexander:2008hd}, a
gauge fixing function which reduces to that of the $R_\xi$ gauge in
the false vacuum has been adopted: ${\cal F}^{(R_\xi)} = \partial_\mu
A_\mu - 2 \xi g (\mbox{Re}\Phi)(\mbox{Im}\Phi)$.  However, such a
gauge fixing function causes a problem when $v=0$ \cite{Endo:2017gal}.
In such a case, the symmetry is restored in the false vacuum and there
appears a class of independent bounce configurations related by the
internal $U(1)$ symmetry.  The configuration is given by
$(A_\mu,\Phi)=(\partial_\mu \Theta /g, \bar{\phi}
e^{i\Theta}/\sqrt{2})$, where the function $\Theta$ obeys
\begin{align}
  \partial_r^2 \Theta + \frac{3}{r} \partial_r \Theta 
  - \frac{1}{2} \xi g^2 \bar{\phi}^2 \sin 2 \Theta = 0.
  \label{eq_Theta}
\end{align}
The above configuration satisfies the classical equation of motion as
well as the boundary condition given in Eq.\ \eqref{bc_bounce}.  The
function $\Theta$ is required to be finite, and is determined by its
value at $r=0$; an independent set of the bounce configurations is
obtained with $0\leq\Theta(0)< 2\pi$.  When $v=0$, we should take
account of all the bounce configurations labeled by $\Theta(0)$ for the
calculation of the decay rate of the false vacuum.  However, it is
highly non-trivial because the fluctuation operators around the bounce
depend on $\Theta (0)$, and also because we have to understand the
measure for the integration over $\Theta (0)$.

Such complications can be avoided with a gauge fixing function which
does not contain the scalar field \cite{Kusenko:1996bv}.  In our
analysis, we adopt the following gauge fixing function:
\begin{align}
  {\cal F} = \partial_\mu A_\mu,
  \label{gaugefixingfn}
\end{align} 
with which the gauge fixing term is given by
\begin{align}
  {\cal L}_{\rm G.F.} = \frac{1}{2\xi} {\cal F}^2.
\end{align}
In addition, the FP ghosts become free fields:
\begin{align}
  {\cal L}_{\rm ghost} =
  - \bar{c} \partial_\mu \partial_\mu c.
\end{align}

With the gauge fixing function given in Eq.\ \eqref{gaugefixingfn}, we
find a class of solution of the classical equation of motion, which is
given by
\begin{align}
  (A_\mu, \Phi)_{\rm classical\ solution} = 
  (0, \bar{\phi} (r) e^{i\vartheta}/\sqrt{2}),
  \label{ClassicalSolution}
\end{align}
with $\vartheta$ being a constant, parameterizing the configurations
of the classical solutions.  When $v\neq 0$, only one configuration
with $\vartheta =0$ satisfies the boundary condition given in Eq.\
\eqref{bc_bounce}.  Then, we can easily integrate out the fluctuations
around the bounce, as we will discuss in Section \ref{sec:nonzerovev}.
On the contrary, when $v=0$, all the classical solutions parameterized
by $\vartheta$ contribute to the false-vacuum decay because all the
bounce parameterized by $\vartheta$ has the same asymptotic value at
$r\rightarrow\infty$, and contributes to the vacuum decay.  This issue
will be discussed in Section \ref{sec:zerovev}.

\subsection{Fluctuation operators}

For the evaluation of the decay rate of the false vacuum, it is
necessary to integrate out the fluctuations around the bounce.  Such
an integration can be performed by calculating the functional
determinants of the second order differential operators (i.e.,
fluctuation operators).  It is convenient to decompose the gauge field
and $\Phi$ as
\begin{align}
  A_\mu = a_\mu,~~~
  \Phi = \frac{1}{\sqrt{2}} e^{i\vartheta}
  \left( \bar{\phi} + h + i \varphi \right),
  \label{Expansion_APhi}
\end{align}
with $h$ and $\varphi$ being real modes.  Hereafter, we call $h$ and
$\varphi$ as ``Higgs'' and ``NG'' modes, respectively.

With the gauge fixing function given in Eq.\ \eqref{gaugefixingfn},
the fluctuation operator of the gauge and NG modes around the bounce
configuration is
\begin{align}
  {\cal M}^{(A_\mu,\varphi)} \equiv &
  \left(
    \begin{array}{cc}
      \displaystyle{ -\partial^2 \delta_{\mu\nu} 
        + \left( 1 - \frac{1}{\xi} \right) \partial_\mu \partial_\nu
        + g^2 \bar{\phi}^2
      }
      & 
      g (\partial_\nu \bar{\phi}) - g \bar{\phi} \partial_\nu
      \\[3mm]
      2 g (\partial_\mu \bar{\phi}) + g \bar{\phi} \partial_\mu
      &
      \displaystyle{ -\partial^2 
        + \frac{(\partial^2 \bar{\phi})}{\bar{\phi}} 
      }
    \end{array}
  \right).
  \label{M_tot}
\end{align}
Here, the derivatives act on everything to the right unless brackets
exist; for example, with the expression $(\partial_\mu \bar{\phi})$,
the derivative acts only on $\bar{\phi}$.  The calculation of the
functional determinant of ${\cal M}^{(A_\mu,\varphi)}$ is the main
subject of this paper.

Because the bounce configuration is $O(4)$ symmetric
\cite{Coleman:1977th, Blum:2016ipp}, the fluctuations can be expanded
by using hyperspherical functions on $S^3$.  The hyperspherical
functions are labeled by the quantum numbers of the rotational group
of the 4D Euclidean space, i.e., $SU(2)_A\times SU(2)_B$ (the so-called
$A$- and $B$-spins).  Namely, we denote ${\cal
  Y}_{J,m_A,m_B}\equiv\langle\hat{\bf r}|J,m_A,m_B\rangle$; $\hat{\bf
  r}$ is the coordinate on $S^3$, the eigenvalue of $S_A^2$ and
$S_B^2$ is $J(J+1)$, and that of $S_{A,3}$ ($S_{B,3}$) is $m_A$
($m_B$), where generators of $SU(2)_A$ and $SU(2)_B$ are denoted as
$S_A$ and $S_B$, respectively.  With introducing radial mode functions
denoted as $\alpha_S$, $\alpha_L$, $\alpha_{T1}$, and $\alpha_{T2}$,
$a_\mu$ can be expanded as
\begin{align}
  a_\mu (x) \ni &\,
  \alpha_S (r) \frac{x_\mu}{r} {\cal Y}_{J,m_A,m_B}
  + \alpha_L (r) \frac{r}{L} \partial_\mu {\cal Y}_{J,m_A,m_B}
  \nonumber \\ &\, 
  + \alpha_{T1} (r) 
  i \epsilon_{\mu\nu\rho\sigma} V^{(1)}_\nu L_{\rho\sigma} {\cal Y}_{J,m_A,m_B}
  + \alpha_{T2} (r) 
  i \epsilon_{\mu\nu\rho\sigma} V^{(2)}_\nu L_{\rho\sigma} {\cal Y}_{J,m_A,m_B},
\end{align}
where $V^{(1)}_\nu$ and $V^{(2)}_\nu$ are (arbitrary) two independent
vectors,
$L_{\rho\sigma}\equiv\frac{i}{\sqrt{2}}
(x_\rho\partial_\sigma-x_\sigma\partial_\rho)$,
and
\begin{align}
  L \equiv \sqrt{4J (J +1)}.
\end{align}
(Notice that $L_{\mu\nu}L_{\mu\nu}{\cal Y}_{J,m_A,m_B}=L^2{\cal
  Y}_{J,m_A,m_B}$.)  Here, we omit subscripts $J$, $m_A$, and $m_B$
from the mode functions for notational simplicity, and summations over
$J$, $m_A$, and $m_B$ are implicit.  There is no $L$- or $T$-mode for
$J=0$. The scalar bosons can be expanded as
\begin{align}
  h (x) \ni \alpha_h (r) {\cal Y}_{J,m_A,m_B},
  \\
  \varphi (x) \ni \alpha_\varphi (r) {\cal Y}_{J,m_A,m_B}.
\end{align}
The behavior of the radial mode functions are governed by the
fluctuation operators.  In the following, we show explicit expressions
of the fluctuation operators for each angular momentum.

The fluctuation operator for $(\alpha_S,\alpha_L,\alpha_\varphi)$ and
that for $(\alpha_{T1},\alpha_{T2})$ are decoupled from each other.
For $J>0$, the fluctuation operator for
$(\alpha_S,\alpha_L,\alpha_\varphi)$ is given by
\begin{align}
  {\cal M}_J^{(S,L,\varphi)} \equiv &
  \left(
    \begin{array}{ccc}
      \displaystyle{ -\Delta_J + \frac{3}{r^2} + g^2 \bar{\phi}^2 }
      & \displaystyle{ -\frac{2L}{r^2} } 
      & g \bar{\phi}' - g \bar{\phi} \partial_r
      \\[3mm]
      \displaystyle{ -\frac{2L}{r^2} } 
      & \displaystyle{ -\Delta_J - \frac{1}{r^2} + g^2 \bar{\phi}^2 }
      & \displaystyle{ -\frac{L}{r} g \bar{\phi} } 
      \\[3mm]
      2 g \bar{\phi}' + g \bar{\phi} \partial_r + 
      \displaystyle{ \frac{3}{r} g \bar{\phi} }
      & \displaystyle{ -\frac{L}{r} g \bar{\phi} } 
      & \displaystyle{ 
        -\Delta_J + \frac{(\Delta_0 \bar{\phi})}{\bar{\phi}} 
      }
    \end{array}
  \right)
  \nonumber \\[2mm] & + 
  \left( 1 - \frac{1}{\xi} \right)
  \left(
    \begin{array}{ccc}
      \displaystyle{ \partial_r^2 + \frac{3}{r} \partial_r - \frac{3}{r^2} }
      & \displaystyle{ 
        -L \left(\frac{1}{r} \partial_r - \frac{1}{r^2} \right) } 
      & 0 
      \\[3mm]
      \displaystyle{ L \left( \frac{1}{r} \partial_r + \frac{3}{r^2} \right) } 
      & \displaystyle{ -\frac{L^2}{r^2} }
      & 0
      \\[3mm]
      0 & 0 & 0
    \end{array}
  \right),  
  \label{M_J}
\end{align}
where $\bar{\phi}'\equiv\partial_r\bar{\phi}$, and
\begin{align}
  \Delta_J \equiv \partial_r^2 + \frac{3}{r} \partial_r 
  - \frac{L^2}{r^2}.
\end{align}
For $J=0$, $\alpha_L$-mode does not exist, and the fluctuation
operator is in the form of $2\times 2$ differential operator; ${\cal
  M}_{J=0}^{(S,\varphi)}$ is obtained from Eq.\ \eqref{M_J} by
eliminating the second row and the second column:
\begin{align}
  {\cal M}_{J=0}^{(S,\varphi)} \equiv &
  \left(
    \begin{array}{cc}
      \displaystyle{ 
        \frac{1}{\xi} \left( 
          -\Delta_0 + \frac{3}{r^2} + \xi g^2 \bar{\phi}^2 
        \right) }
      & g \bar{\phi}' - g \bar{\phi} \partial_r
      \\[3mm]
      2 g \bar{\phi}' + g \bar{\phi} \partial_r + 
      \displaystyle{ \frac{3}{r} g \bar{\phi} }
      &
      \displaystyle{ 
        -\Delta_0 + \frac{(\Delta_0 \bar{\phi})}{\bar{\phi}}
      }
    \end{array}
  \right).
  \label{M_J=0}
\end{align}
In addition, the fluctuation operator of the transverse modes is
given by
\begin{align}
  {\cal M}_J^{(T)} = -\Delta_J + g^2 \bar{\phi}^2.
  \label{M^T}
\end{align}
while that of the FP ghosts is
\begin{align}
  {\cal M}_J^{(\bar{c},c)} = -\Delta_J.
  \label{MFP}
\end{align}
The radial mode functions can be expanded by using the eigenfunctions
of these fluctuation operators.

We also need fluctuation operators around the false vacuum, which are
denoted as $\widehat{\cal M}^{(A_\mu,\varphi)}$, $\widehat{\cal
  M}_J^{(S,L,\varphi)}$, $\widehat{\cal M}_J^{(T)}$, and so on.  (In
this paper, the ``hat'' is used for objects around the false vacuum.)
They can be obtained from the corresponding fluctuation operators
around the bounce by replacing $\bar{\phi}\rightarrow v$, and
$\bar{\phi}'\rightarrow 0$.  For the case of $v=0$,
$(\Delta_0\bar{\phi})/\bar{\phi}$ should be replaced by $m_\phi^2$.

Finally, let us comment on the contribution of the Higgs mode.  In
this paper, we concentrate on the case where there exists only one
charged scalar field which has non-vanishing amplitude.  However,
assuming renormalizability, extra neutral scalars are necessary to
make the scalar potential to have both false and true vacua; we
implicitly assume that this is the case.  The neutral scalar fields
may mix with the charged scalar when the $U(1)$ symmetry is broken.
Thus, the fluctuation operators of the CP-even scalars (which includes
the Higgs mode) are highly model dependent.\footnote
{In our study, we assume no significant CP violation in the scalar
  sector, i.e., no mixing between CP-even and CP-odd scalars.}
However, the fluctuation operators of the CP-even scalars do not
depend on $\xi$, and hence have nothing to do with the gauge
dependence of the decay rate of the false vacuum, which is of our
primary concern.  Therefore, we do not discuss the effects of the
Higgs mode.

\subsection{Prefactor ${\cal A}$}

The prefactor ${\cal A}$ in Eq.\ \eqref{decayrate} is given by
\cite{Callan:1977pt}
\begin{align}
  {\cal A} = \frac{{\cal B}^2}{4\pi^2} 
  {\cal A}'^{(h)} 
  {\cal A}^{(A_\mu,\varphi)} 
  {\cal A}^{(\bar{c},c)} 
  {\cal A}^{(\rm extra)}
  e^{-{\cal S}^{\rm (c.t.)}},
  \label{prefactorA}
\end{align}
where ${\cal A}'^{(h)}$, ${\cal A}^{(A_\mu,\varphi)}$, and ${\cal
  A}^{(\bar{c},c)}$ are contributions of the Higgs mode,
$(A_\mu,\varphi)$, and the FP ghosts, respectively.  The ``prime'' on
${\cal A}'^{(h)}$ indicates that the effect of the zero modes in
association with the translational invariance is omitted
\cite{Callan:1977pt}.  If there exist extra fields other than those
mentioned above, their contribution is denoted as ${\cal A}^{(\rm
  extra)}$; hereafter, we do not consider ${\cal A}^{(\rm extra)}$.
In addition, ${\cal S}^{\rm (c.t.)}$ is the counter term to subtract
the divergences.

At the one-loop level, each contribution is obtained by evaluating the
functional determinants of the fluctuation operators.  In particular,
formally, ${\cal A}^{(A_\mu,\varphi)}$ is given in the following form
\begin{align}
  {\cal A}^{(A_\mu,\varphi)} = 
  \left[
    \frac{\mbox{Det} {\cal M}^{(A_\mu,\varphi)}}
    {\mbox{Det} \widehat{\cal M}^{(A_\mu,\varphi)}}
  \right]^{-1/2}.
\end{align}
It can be further decomposed into the contributions of
$(\alpha_S,\alpha_L,\varphi)$ and $(\alpha_{T1},\alpha_{T2})$, which
are denoted as ${\cal A}^{(S,L,\varphi)}$ and ${\cal A}^{(T)}$,
respectively, as
\begin{align}
  {\cal A}^{(A_\mu,\varphi)} = {\cal A}^{(S, L, \varphi)} {\cal A}^{(T)},
\end{align}
where 
\begin{align}
  {\cal A}^{(S,L,\varphi)} = &\,
  \left[
    \frac{\mbox{Det} {\cal M}_0^{(S,\varphi)}}
    {\mbox{Det} \widehat{\cal M}_0^{(S,\varphi)}}
  \right]^{-1/2}
  \prod_{J=1/2}^{\infty} 
  \left[
    \frac{\mbox{Det} {\cal M}_J^{(S,L,\varphi)}}
    {\mbox{Det} \widehat{\cal M}_J^{(S,L,\varphi)}}
  \right]^{-(2J +1)^2/2},
  \\
  {\cal A}^{(T)} = &\,
  \prod_{J=1/2}^{\infty} 
  \left[
    \frac{\mbox{Det} {\cal M}_J^{(T)}}
    {\mbox{Det} \widehat{\cal M}_J^{(T)}}
  \right]^{-(2J +1)^2}.
\end{align}
Furthermore, the ghost contribution is given by
\begin{align}
  {\cal A}^{(\bar{c},c)} = 
  \prod_{J=0}^{\infty} 
  \left[
    \frac{\mbox{Det} {\cal M}_J^{(\bar{c},c)}}
    {\mbox{Det} \widehat{\cal M}_J^{(\bar{c},c)}}
  \right]^{(2J +1)^2}.
\end{align}

\subsection{Functional determinant}

In order to evaluate the functional determinants, we use the method
given in \cite{Coleman:aspectsof, Dashen:1974ci, Kirsten:2003py,
  Kirsten:2004qv}.  For fluctuation operators ${\cal M}_J^{(X)}$ and
$\widehat{\cal M}_J^{(X)}$, which are $N\times N$ differential
operators in general, we first consider $N$ independent functions
$\psi_I^{(X)} (r)$ and $\widehat{\psi}_I^{(X)}(r)$ (with $I=1-N$),
which obey
\begin{align}
  &
  {\cal M}_J^{(X)}\psi_I^{(X)}=0,
  \label{generalpsi}
  \\ &
  \widehat{\cal M}_J^{(X)}\widehat{\psi}_I^{(X)}=0.
\end{align}
In addition, $\psi_I^{(X)}$ and $\widehat{\psi}_I^{(X)}$ are regular
at $r\rightarrow 0$.  With these functions, we introduce the following
quantities:
\begin{align}
  &
  {\cal D}^{(X)}_J (r) \equiv 
  \mbox{det} (\psi_1^{(X)}(r)~\cdots~\psi_N^{(X)}(r)),
  \\
  &
  \widehat{\cal D}^{(X)}_J (r) \equiv 
  \mbox{det} (\widehat\psi_1^{(X)}(r)~\cdots~\widehat\psi_N^{(X)}(r)).
\end{align}

When $\psi_I^{(X)} (r)$ and $\widehat{\psi}_I^{(X)}(r)$ have the same
boundary condition at $r\rightarrow 0$, the quantity $\mbox{Det} {\cal
  M}_J^{(X)}/\mbox{Det} \widehat{\cal M}_J^{(X)}$ is given by the
ratio of ${\cal D}_J^{(X)}(r\rightarrow\infty)$ and $\widehat{\cal
  D}_J^{(X)}(r\rightarrow\infty)$.  For a general boundary condition
at $r=0$, the ratio of the functional determinants is given by
\begin{align}
  \frac{\mbox{Det} {\cal M}_J^{(X)}}
  {\mbox{Det} \widehat{\cal M}_J^{(X)}} = 
  \left[
    \frac{{\cal D}_J^{(X)}(r\rightarrow 0)}
    {\widehat{\cal D}_J^{(X)}(r\rightarrow 0)}
  \right]^{-1}
  \frac{{\cal D}_J^{(X)}(r_\infty)}
  {\widehat{\cal D}_J^{(X)}(r_\infty)},
  \label{Gelfand-Yaglom}
\end{align}
where $r_\infty$ is the abbreviation of $r\rightarrow\infty$.  This
relation is derived in Appendix \ref{app:funcdet}.

\section{Useful Formulae}
\label{sec:formulae}
\setcounter{equation}{0}

In this section, we summarize properties of the functions used in our
calculations of the functional determinants in later sections.  The
formulae given in this section are applicable both for $v\neq 0$ and
$v=0$.

\subsection{FP ghosts and transverse modes}

The fluctuation operator for the FP ghosts is given in Eq.\
\eqref{MFP}.  We define the function $f^{\rm (FP)}_J$ which obeys
\begin{align}
  \Delta_J f^{\rm (FP)}_J = 0,
  \label{eq_fFP}
\end{align}
and $f^{\rm (FP)}_J (0)$ is required to be finite; we normalize
$f^{\rm (FP)}_J$ as
\begin{align}
  f^{\rm (FP)}_J (r) = r^{2J}.
\end{align}
(For $J=0$, $f^{\rm (FP)}_0 (r) = 1$.)  The differential equation
$\widehat{\cal M}^{(\bar{c},c)}_J \widehat{f}^{\rm (FP)}_J = 0$ has
the same solution:
\begin{align}
  \widehat{f}^{\rm (FP)}_J (r) = f^{\rm (FP)}_J (r) .
\end{align}
Using Eq.\ \eqref{Gelfand-Yaglom},
\begin{align}
  \frac{\mbox{Det} {\cal M}_J^{(\bar{c},c)}}
  {\mbox{Det} \widehat{\cal M}_J^{(\bar{c},c)}}
  = 1.
\end{align}

For the functional determinants for the transverse modes, we define
the functions $f^{(T)}_J$ and $\widehat{f}^{(T)}_J$, obeying
\begin{align}
  &(\Delta_J - g^2 \bar{\phi}^2) f^{(T)}_J = 0,
  \\
  &(\Delta_J - g^2 v^2) \widehat{f}^{(T)}_J = 0,
\end{align}
with 
\begin{align}
  f^{(T)}_J (r\rightarrow 0) \simeq 
  \widehat{f}^{(T)}_J (r\rightarrow 0) \simeq 
  r^{2J}.
\end{align}
Then,
\begin{align}
  \frac{\mbox{Det} {\cal M}_J^{(T)}}
  {\mbox{Det} \widehat{\cal M}_J^{(T)}}
  =
  \frac{f^{(T)}_J (r_\infty)}
  {\widehat{f}^{(T)}_J (r_\infty)}.
\end{align}

\subsection{$S$, $L$, and NG modes with $J\neq 0$}

In this subsection, we consider the functional determinants of the
fluctuation operators for the $S$, $L$, and NG modes with $J\neq 0$.
For this purpose, we should consider the equations
\begin{align}
  {\cal M}^{(S,L,\varphi)}_J \Psi = 0,
  \label{Ml*Psi=0}
\end{align}
and
\begin{align}
  \widehat{\cal M}^{(S,L,\varphi)}_J \widehat{\Psi} = 0.
  \label{hatMl*Psi=0}
\end{align}

First, we consider Eq.\ \eqref{Ml*Psi=0}.  It is convenient to use the
fact that the solutions of Eq.\ \eqref{Ml*Psi=0} can be expressed by
using three functions which we call $\chi$, $\eta$, and $\zeta$.  Let
us define
\begin{align}
  \Psi \equiv
  \left( \begin{array}{c}
      \Psi^{\rm (top)} \\
      \Psi^{\rm (mid)} \\
      \Psi^{\rm (bot)}
    \end{array} 
  \right) 
  \equiv
  \left( \begin{array}{c}
      \partial_r \chi 
      \\[3mm]
      \displaystyle{\frac{L}{r}} \chi
      \\[3mm] 
      g \bar{\phi} \chi
    \end{array} 
  \right) 
  + 
  \left( \begin{array}{c}
      \displaystyle{ \frac{1}{r g^2 \bar{\phi}^2} \eta }
      \\[3mm]
      \displaystyle{ \frac{1}{L r^2 g^2 \bar{\phi}^2} \partial_r (r^2 \eta) }
      \\[3mm]
      0
    \end{array} 
  \right)
  + 
  \left( \begin{array}{c}
      \displaystyle{
        -2 \frac{\bar{\phi}'}{g^2 \bar{\phi}^3} \zeta 
      }
      \\[3mm]
      0
      \\[3mm]
      \displaystyle{
        \frac{1}{g \bar{\phi}} \zeta 
      }
    \end{array} 
  \right),
  \label{Psi}
\end{align}
where the functions $\chi$, $\eta$, and $\zeta$ obey the following
equations:
\begin{align}
  &
  \Delta_J \chi = 
  \frac{2 \bar{\phi}'}{r g^2 \bar{\phi}^3} \eta 
  + \frac{2}{r^3} \partial_r 
  \left( \frac{r^3 \bar{\phi}'}{g^2 \bar{\phi}^3} \zeta \right)
  - \xi \zeta,
  \label{eq_chi}
  \\[2mm] &
  (\Delta_J - g^2 \bar{\phi}^2) \eta 
  - \frac{2 \bar{\phi}'}{r^2 \bar{\phi}} \partial_r
  \left(
    r^2 \eta
  \right)
  = - \frac{2 L^2 \bar{\phi}'}{r \bar{\phi}} \zeta,
  \label{eq_eta}
  \\[2mm] &
  \Delta_J \zeta = 0.
  \label{eq_zeta}
\end{align}
Then, using the equation for the bounce solution given in Eq.\
\eqref{ClassicalEq}, we can show that $\Psi$ satisfies Eq.\
\eqref{Ml*Psi=0}.  Thus, the function $\Psi$ given in Eq.\ \eqref{Psi}
has the right property to calculate the functional determinant of
${\cal M}^{(S,L,\varphi)}_J$, assuming that $\Psi(r=0)$ is finite.  In
addition, the following equations hold for the top and middle
components of $\Psi$, which are also useful for the following
argument:
\begin{align}
  \partial_r \Psi^{\rm (top)} &\, = 
  - \frac{3}{r} \Psi^{\rm (top)}
  + \frac{L}{r} \Psi^{\rm (mid)} - \xi \zeta,
  \label{Psitop'}
  \\
  \partial_r \Psi^{\rm (mid)} &\, = 
  \frac{L}{r} \Psi^{\rm (top)}
  - \frac{1}{r} \Psi^{\rm (mid)} + \frac{1}{L} \eta.
  \label{Psimid'}
\end{align}
Notice that Eq.\ \eqref{Psitop'} can be translated to $\alpha_{\cal
  F}+\xi\zeta=0$, where $\alpha_{\cal F}$ is the radial mode function
of the gauge fixing function, i.e., ${\cal F}(x)\ni\alpha_{\cal
  F}(r){\cal Y}_{J,m_A,m_B}$.

If three independent solutions of Eq.\ \eqref{Ml*Psi=0} are given
(which we call $\Psi_I$ with $I=1$, $2$, and $3$), the functional
determinant of ${\cal M}^{(S,L,\varphi)}_J$ can be related to the
function ${\cal D}^{(S,L,\varphi)}_J$ defined as
\begin{align}
  {\cal D}^{(S,L,\varphi)}_J (r)
  \equiv 
  \mbox{det} (\Psi_1 (r) ~ \Psi_2 (r) ~ \Psi_3 (r)).
  \label{det}
\end{align}
We consider the following three independent solutions, which are
composed of the functions $\chi_I$, $\eta_I$, and $\zeta_I$.
\begin{enumerate}
\item We take $\eta_1=\zeta_1=0$; this condition is consistent with
  Eqs.\ \eqref{eq_eta} and \eqref{eq_zeta}.  Then, Eq.\ \eqref{eq_chi}
  can be easily solved to obtain $\chi_1$ by requiring its regularity
  at the origin; we normalize $\chi_1$ as
  \begin{align}
    \chi_1 (r) =  r^{2J},
  \end{align}
  which gives
  \begin{align}
    \Psi_1 (r) = 
     \left( \begin{array}{c}
         2J r^{2J-1} \\ L r^{2J-1} \\ g \bar{\phi} r^{2J}
       \end{array}
     \right).
     \label{BC:Psi1}
  \end{align}
\item We take $\zeta_2=0$ (which is consistent with \eqref{eq_zeta}),
  while $\chi_2$ and $\eta_2$ are non-vanishing.  For the boundary
  conditions at $r\rightarrow 0$, we take
  \begin{align}
    \chi_2 (r\rightarrow 0) &\, \simeq -
    \frac{1}{2J g^2 \bar{\phi}_C^2} r^{2J},
    \\
    \eta_2 (r\rightarrow 0) &\, \simeq r^{2J},
  \end{align}
  where $\bar{\phi}_C$ is the scalar amplitude at the center of the
  bounce:
  \begin{align}
    \bar{\phi}_C \equiv \bar{\phi} (r=0).
  \end{align}
  We find
  \begin{align}
    \Psi_2 (r\rightarrow 0) \simeq 
     \left( \begin{array}{c}
         \displaystyle{ \frac{1}{8(J+1)} r^{2J+1} }
         \\[3mm]
         \displaystyle{ \frac{J (J+2)}{L^3} r^{2J+1} }
         \\[3mm]
         \displaystyle{ -\frac{1}{2J g \bar{\phi}_C} r^{2J} }
       \end{array}
     \right).
     \label{BC:Psi2}
  \end{align}
\item For $\Psi_3$, we take
  \begin{align}
    \zeta_3 (r) = r^{2J},
  \end{align}
  and $\chi_3 (r\rightarrow 0)\sim \eta_3 (r\rightarrow 0)\sim
  O(r^{2J+2})$.  Then, we find
  \begin{align}
    \Psi_3 (r\rightarrow 0) \simeq 
     \left( \begin{array}{c}
         \displaystyle{ -\frac{1}{4} \xi r^{2J + 1} }
         \\[3mm] 
         \displaystyle{ - \frac{J}{2L} \xi r^{2J + 1} }
         \\[3mm] 
         \displaystyle{ \frac{1}{g \bar{\phi}_C} r^{2J} }
       \end{array}
     \right).
     \label{BC:Psi3}
  \end{align}
\end{enumerate}
Combining Eqs.\ \eqref{BC:Psi1}, \eqref{BC:Psi2}, and \eqref{BC:Psi3},
we obtain
\begin{align}
  {\cal D}_J^{(S,L,\varphi)} (r\rightarrow 0) \simeq 
  \frac{2J [(J+1) \xi + J]}{L^3 g \bar{\phi}_C}
  r^{6J}.
  \label{BC:D_J(S,L,NG)}
\end{align}

In order to express the homogeneous solution of Eq.\ \eqref{eq_eta}
with $\zeta=0$, we introduce the function $f^{(\eta)}_J$, which obeys
\begin{align}
  (\Delta_J - g^2 \bar{\phi}^2) f^{(\eta)}_J
  - \frac{2 \bar{\phi}'}{r^2 \bar{\phi}} \partial_r
  \left(
    r^2 f^{(\eta)}_J
  \right)
  = 0,
  \label{eq_feta}
\end{align}
whose boundary condition is taken to be
\begin{align}
  f^{(\eta)}_J (r\rightarrow 0) \simeq r^{2J}.
\end{align}
(For $J=0$, $f^{(\eta)}_0 (r\rightarrow 0) \simeq 1$.)  At
$r\rightarrow\infty$, $f^{(\eta)}_J$ behaves as
\begin{align}
  f^{(\eta)}_J (r\rightarrow\infty) &\, \simeq c_\eta
  \frac{e^{gv r}}{r^{3/2}}
  \left[ 1 + O(r^{-1}) \right],
\end{align}
where $c_\eta$ is a constant.  We emphasize here that the function
$f^{(\eta)}_J$ is independent of $\xi$.  The homogeneous solutions of
Eqs.\ \eqref{eq_chi} and \eqref{eq_zeta} (that of Eq.\ \eqref{eq_eta})
are given by $f^{\rm (FP)}_J$ ($f^{(\eta)}_J$).

For the case of $v\neq 0$, we also define the function
$\widehat{f}^{(\eta)}_J$, obeying
\begin{align}
  (\Delta_J - g^2 v^2) \widehat{f}^{(\eta)}_J = 0,
\end{align}
and
\begin{align}
  \widehat{f}^{(\eta)}_J (r\rightarrow 0) \simeq r^{2J}.
\end{align}
The function $\widehat{f}^{(\eta)}_J$ is given by
\begin{align}
  \widehat{f}^{(\eta)}_J (r) = 2^{2J + 1} \Gamma (2 J + 2)
  (gv)^{-(2J +1)}
  \frac{I_{2J +1} (g v r)}{r},
\end{align}
where $I_{2J +1}$ is the modified Bessel function.  Notice that, when
$v=0$, we will not use the function $\widehat{f}^{(\eta)}_J$ and hence
is not defined.

\subsection{$S$ and NG modes with $J=0$}

In this subsection, we consider the $S$ and NG modes with $J=0$.  The
fluctuation operator ${\cal M}_{J=0}^{(S,\varphi)}$ is in $2\times 2$
form, and the solutions of the equation,
\begin{align}
  {\cal M}_{J=0}^{(S,\varphi)}\Psi=0,
  \label{M0*Psi=0}
\end{align}
can be written as follows:
\begin{align}
  \Psi \equiv
  \left( \begin{array}{c}
      \Psi^{\rm (top)} \\
      \Psi^{\rm (bot)}
    \end{array} 
  \right) 
  \equiv
  \left( \begin{array}{c}
      \partial_r \chi 
      \\
      g \bar{\phi} \chi
    \end{array} 
  \right) 
  + 
  \left( \begin{array}{c}
      \displaystyle{
        -2 \frac{\bar{\phi}'}{g^2 \bar{\phi}^3} \zeta 
      }
      \\[3mm] 
      \displaystyle{ \frac{1}{g \bar{\phi}} \zeta }
    \end{array} 
  \right),
  \label{Psi_0}
\end{align}
where the functions $\chi$ and $\zeta$ obey Eq.\ \eqref{eq_chi} with
$\eta=0$ and Eq.\ \eqref{eq_zeta}, respectively.

As two independent solutions, we adopt the followings:
\begin{enumerate}
\item We take $\zeta_1=0$, and
  \begin{align}
    \chi_1 (r) = 1,
  \end{align}
  which gives 
  \begin{align}
    \Psi_1 (r) =
    \left( \begin{array}{c}
        0 
        \\
        g \bar{\phi}
      \end{array}
    \right).
  \end{align}
\item We take
  \begin{align}
    \zeta_2 (r) = 1,
  \end{align}
  while $\chi_2 (r\rightarrow 0)\sim O(r^2)$.  Then,
  \begin{align}
    \Psi_2 (r\rightarrow 0) \simeq 
    \left( \begin{array}{c}
        \displaystyle{ - \frac{\xi}{4} r }
        \\[3mm]
        \displaystyle{ \frac{1}{g\bar{\phi}_C} }
      \end{array}
    \right).
  \end{align}
\end{enumerate}
Consequently, we find
\begin{align}
  {\cal D}_0^{(S,\varphi)} (r\rightarrow 0) \simeq 
  \frac{1}{4} \xi g \bar{\phi}_C r.
  \label{BC:D_0(S,NG)}
\end{align}

\section{Functional Determinants: Case with $v\neq 0$}
\label{sec:nonzerovev}
\setcounter{equation}{0}

Now, we are at the position to discuss the decay rate of the false
vacuum.  In this section, we consider the case with $v\neq 0$.
Choosing the false vacuum as $(A_\mu,\Phi)_{\rm false\,
  vacuum}=(0,v/\sqrt{2})$, the bounce solution is uniquely determined:
\begin{align}
  (A_\mu,\Phi)_{\rm bounce}=(0,\bar{\phi}/\sqrt{2}).
\end{align}

As we show in Eq.\ \eqref{prefactorA}, the prefactor ${\cal A}$ is
given by the product of the functional determinants of the fluctuation
operators.  In the following, we discuss each contribution separately.

\subsection{$v\neq 0$: contribution of $J\neq 0$}

Let us discuss the behavior of the functions $\Psi_I$ at
$r\rightarrow\infty$.  The behavior of $\chi_I$, $\eta_I$, and
$\zeta_I$ can be understood by using the fact that $\bar{\phi}'$ is
exponentially suppressed at $r\rightarrow\infty$.
\begin{enumerate}
\item Because $\chi_1 (r) = r^{2J}$,
  \begin{align}
    \Psi_1 (r\rightarrow\infty) \simeq
    \left( \begin{array}{c}
        2J r^{2J-1} 
        \\ 
        L r^{2J-1} 
        \\ g v r^{2J}
      \end{array}
    \right).
  \end{align}
\item Because $\chi_2$ is given by the sum of a homogeneous solution
  and a particular solution (which we denote $\delta\chi^{(\eta)}$),
  the second set of the mode functions can be expressed as
  \begin{align}
    \chi_2 (r) &\, = 
    a_1 r^{2J} + \delta\chi^{(\eta)} (r),
    \label{chi2(vneq0)}
    \\
    \eta_2 (r) &\, = 
    f^{(\eta)}_J (r),
  \end{align}
  where $a_1$ is a constant.  (Here and hereafter, dumping modes are
  neglected.)  The function $\delta\chi^{(\eta)}$ satisfies the
  following equation:
  \begin{align}
    \Delta_J \delta\chi^{(\eta)} = 
    \frac{2 \bar{\phi}'}{r g^2 \bar{\phi}^3} f^{(\eta)}_J.
  \end{align}
  We eliminate a term proportional to $r^{2J}$ in
  $\delta\chi^{(\eta)}$ with the redefinition of $a_1$.  Then, at
  $r\rightarrow\infty$, $\delta\chi^{(\eta)}$ behaves as
  \begin{align}
    \delta\chi^{(\eta)} (r\rightarrow\infty ) \simeq
    - \frac{2 m_\phi \kappa}{g^2 v^3 (gv-m_\phi)^2 r^{5/2}}
    e^{-m_\phi r} f^{(\eta)}_J
    + \cdots.
  \end{align}
  (In Appendix \ref{app:asymptotic}, we discuss a procedure to derive
  asymptotic behavior of the solutions of the differential equation of
  this type.)  Thus, the asymptotic behavior of $\Psi_2$ is given by
  \begin{align}
    \Psi_2 (r\rightarrow\infty) \simeq
    a_1 \Psi_1 (r\rightarrow\infty) + 
    \left( \begin{array}{c}
        O(r^{-1} f_J^{(\eta)})
        \\[2mm]
        \displaystyle{
          \frac{1}{L g^2 v^2} \partial_r f_J^{(\eta)}
        }
        \\[2mm]
        O(r^{-5/2} e^{-m_\phi r} f_J^{(\eta)})
      \end{array}
    \right).
  \end{align}
\item For $\Psi_3$, we denote
  \begin{align}
    \chi_3 (r) &\, =
    b_1 r^{2J}  + b_2 \delta\chi^{(\eta)} (r) + 
    \delta\chi^{(\zeta)} (r),
    \label{chi_3}
    \\
    \eta_3 (r) &\, =
    b_2 f^{(\eta)}_J (r) + \delta\eta^{(\zeta)} (r),
    \label{eta_3}
    \\
    \zeta_3 (r) &\, = r^{2J},
  \end{align}
  with $b_1$ and $b_2$ being constants.  The functions
  $\delta\chi^{(\zeta)}$ and $\delta\eta^{(\zeta)}$ obey the following
  equations:
  \begin{align}
    &
    \Delta_J \delta\chi^{(\zeta)} = 
    \frac{2 \bar{\phi}'}{r g^2 \bar{\phi}^3} \delta\eta^{(\zeta)}
    + \frac{2}{r^3} \partial_r 
    \left( \frac{\bar{\phi}'}{g^2 \bar{\phi}^3} r^{2J +3} \right)
    - \xi r^{2J},
    \\[2mm] &
    (\Delta_J - g^2 \bar{\phi}^2) \delta\eta^{(\zeta)}
    - \frac{2 \bar{\phi}'}{r^2 \bar{\phi}} \partial_r
    \left(
      r^2 \delta\eta^{(\zeta)}
    \right)
    = - \frac{2 L^2 \bar{\phi}'}{\bar{\phi}}  r^{2J -1},
    \label{Eqfordeleta}
  \end{align}
  and they asymptotically behave as
  \begin{align}
    \delta\chi^{(\zeta)} (r\rightarrow\infty) &\, \simeq
    -\frac{1}{8(J +1)} \xi r^{2J +2},
    \\
    \delta\eta^{(\zeta)} (r\rightarrow\infty) &\, \simeq  0.
  \end{align}
  Then, we obtain
  \begin{align}
    \Psi_3 (r\rightarrow\infty) \simeq
    b_1 \Psi_1 (r\rightarrow\infty) + 
    b_2 \Psi_2 (r\rightarrow\infty) 
    - \frac{\xi}{8(J +1)}
    \left( \begin{array}{c}
        (2J + 2) r^{2J +1}
        \\
        L r^{2J +1}
        \\
        g v r^{2J +2}
      \end{array}
    \right).
  \end{align}
\end{enumerate}
Consequently, the determinant defined in Eq.\ \eqref{det} behaves as
\begin{align}
  {\cal D}^{(S,L,\varphi)}_J (r\rightarrow\infty) \sim
  \mbox{det}
  \left( \begin{array}{ccc}
      O(r^{2J -1}) &
      O(r^{-1} f^{(\eta)}_J) &
      O(r^{2J +1}) 
      \\
      O(r^{2J -1}) &
      O(f^{(\eta)}_J) &
      O(r^{2J +1}) 
      \\
      O(r^{2J}) &
      O(r^{-5/2} e^{-m_\phi r} f^{(\eta)}_J) &
      O(r^{2J +2}) 
    \end{array}
  \right),
\end{align}
and
\begin{align}
  {\cal D}^{(S,L,\varphi)}_J (r\rightarrow\infty) \simeq
  \frac{J \xi}{L^3} f^{(\eta)}_J r^{4J+1}.
\end{align}

For the calculation of the decay rate of the false vacuum, we also
need the functional determinants around the false vacuum.  Notice
that, in the argument so far, we have only used the fact that
$\bar{\phi}$ is a solution of the classical equation of motion and is
non-vanishing.  Thus, three independent solutions of Eq.\
\eqref{hatMl*Psi=0} can be obtained by the same argument but replacing
$\bar{\phi}\rightarrow v$.  (Notice that $\Phi=v/\sqrt{2}$ is also a
solution of the classical equation of motion.)  We find
\begin{align}
  \frac{{\cal D}^{(S,L,\varphi)}_J (r\rightarrow 0)}
  {\widehat{\cal D}^{(S,L,\varphi)}_J (r\rightarrow 0)} = &\,
  \frac{v}{\bar{\phi}_C},
\end{align}
and 
\begin{align}
  \frac{{\cal D}^{(S,L,\varphi)}_J (r\rightarrow \infty)}
  {\widehat{\cal D}^{(S,L,\varphi)}_J (r\rightarrow \infty)} = &\,
  \frac{f_J^{(\eta)}}
  {\widehat{f}_J^{(\eta)}},
\end{align}
leading to
\begin{align}
  \frac{\mbox{Det} {\cal M}_J^{(S,L,\varphi)}}
  {\mbox{Det} \widehat{\cal M}_J^{(S,L,\varphi)}}
  = 
  \frac{\bar{\phi}_C f_J^{(\eta)}(r_\infty)}
  {v \widehat{f}_J^{(\eta)}(r_\infty)}.
\end{align}

\subsection{$v\neq 0$: contribution of $J= 0$}

The case of $J=0$ is similar to that for $J\neq 0$.  Thus, we just
show the results.  Using Eq.\ \eqref{BC:D_0(S,NG)},
\begin{align}
  \frac{{\cal D}^{(S,\varphi)}_0 (r\rightarrow 0)}
  {\widehat{\cal D}^{(S,\varphi)}_0 (r\rightarrow 0)} = &\,
  \frac{\bar{\phi}_C}{v}.
\end{align}
In addition, the asymptotic behavior of $\Psi_I$ (with $I=1$ and $2$)
is
\begin{align}
  \Psi_1 (r\rightarrow \infty) \simeq &\,
  \left( \begin{array}{c}
      0
      \\
      g v
    \end{array}
  \right),
  \\[3mm]
  \Psi_2 (r\rightarrow \infty) \simeq &\,
  b_1 \Psi_1 (r\rightarrow \infty) 
  - \frac{1}{8} \xi 
  \left( \begin{array}{c}
      2 r
      \\
      g v r^2
    \end{array}
  \right),
\end{align}
with $b_1$ being a constant, and hence
\begin{align}
  \frac{{\cal D}^{(S,L,\varphi)}_J (r\rightarrow \infty)}
  {\widehat{\cal D}^{(S,L,\varphi)}_J (r\rightarrow \infty)} = 1.
\end{align}
Consequently,
\begin{align}
  \frac{\mbox{Det} {\cal M}_0^{(S,\varphi)}}
  {\mbox{Det} \widehat{\cal M}_0^{(S,\varphi)}}
  = \frac{v}{\bar{\phi}_C}.
\end{align}

\subsection{$v\neq 0$: final result}

Combining the contributions of $J=0$ and $J\neq 0$, we obtain
\begin{align}
  {\cal A}^{(S,L,\varphi)}
  = 
  \left(
    \frac{\bar{\phi}_C}{v}
  \right)^{1/2}
  \prod_{J\geq 1/2}
  \left[
    \frac{\bar{\phi}_C f^{(\eta)}_J (r_\infty)}
    {v \widehat{f}^{(\eta)}_J (r_\infty)}
  \right]^{-(2J +1)^2/2}.
  \label{A^SLphi(final)}
\end{align}
The above expression is manifestly $\xi$-independent.  In addition,
the other contributions are given by
\begin{align}
  {\cal A}^{(T)} =&\,
  \prod_{J=1/2}^{\infty} \left[
    \frac{f^{(T)}_J (r_\infty)}{\widehat{f}^{(T)}_J (r_\infty)}
  \right]^{-(2J +1)^2},
  \label{A^T(final)}
  \\
  {\cal A}^{(\bar{c},c)} =&\,
  1.
\end{align}
When $v\neq 0$, $\widehat{f}^{(\eta)}_J
(r_\infty)\simeq\widehat{f}^{(\infty)}_J (r;gv)$, and
$\widehat{f}^{(T)}_J (r_\infty)\simeq\widehat{f}^{(\infty)}_J (r;gv)$,
where
\begin{align}
  \widehat{f}^{(\infty)}_J (r;m) \equiv 
  \sqrt{\frac{2}{\pi}} \Gamma (2J+2) 
  \left( \frac{m}{2} \right)^{-2J}
  \frac{e^{mr}}{(mr)^{3/2}}.
\end{align}
Notice that the contribution of each angular momentum $J$ is finite.

The above result can be compared with that with another gauge fixing
function, ${\cal F}^{(R_\xi)} = \partial_\mu A_\mu - 2 \xi g
(\mbox{Re}\Phi) (\mbox{Im}\Phi)$, which has been used in literature.
For $v\neq 0$, the calculation with ${\cal F}^{(R_\xi)}$ has been
performed in \cite{Endo:2017gal}.  We can see that the two analyses
give the same expression for the decay rate of the false vacuum.

\section{Functional Determinants: Case with $v= 0$}
\label{sec:zerovev}
\setcounter{equation}{0}

Next, we discuss the case in which the $U(1)$ symmetry is unbroken in
the false vacuum, i.e., $v=0$.  As discussed in Section
\ref{sec:setup}, when $v=0$, there exists a class of bounce (i.e., the
solution of the classical equation of motion with relevant boundary
condition).  All the bounce configurations are related by the global
$U(1)$ transformation which is unbroken at the false vacuum.  We
parameterize the bounce configurations for the gauge fixing function
given in Eq.\ \eqref{gaugefixingfn} as
\begin{align}
  (A_\mu, \Phi)_{\rm bounce} = 
  (0, \bar{\phi} e^{i\vartheta}/\sqrt{2}),
  \label{bounce_v=0}
\end{align}
where $0\leq\vartheta <2\pi$.  We need to take account of all the
bounce configurations in the calculation of the decay rate, as we will
discuss below, which requires the integration over the variable
$\vartheta$.

Importantly, expanding the gauge and scalar fields as Eq.\
\eqref{Expansion_APhi}, the fluctuation operators are
$\vartheta$-independent.  This is an advantage of taking the gauge
fixing function given in Eq.\ \eqref{gaugefixingfn}, with which the
integration over $\vartheta$ is easily performed.

\subsection{$v=0$: contribution of $J\neq 0$}

In this subsection, we consider the contribution of $J\neq 0$ modes to
${\cal A}^{(S,L,\varphi)}$.  Even for the case of $v=0$, Eqs.\
\eqref{Psi} $-$ \eqref{eq_zeta} are useful to study the functional
determinants.

The behavior of the function $\Psi_I$ around $r\rightarrow 0$ is
insensitive to the value of $v$; thus, ${\cal D}_J^{(S,L,\varphi)}
(r\rightarrow 0)$ is given in Eq.\ \eqref{BC:D_J(S,L,NG)} even for
$v=0$.  Next, we consider the asymptotic behavior of ${\cal
  D}_J^{(S,L,\varphi)}(r\rightarrow\infty)$.
\begin{enumerate}
\item For the first solution $\Psi_1$, we are taking
  \begin{align}
    \chi_1 (r\rightarrow\infty) = r^{2J},
  \end{align}
  and hence
  \begin{align}
    \Psi_1 (r\rightarrow \infty) \simeq 
    \left( \begin{array}{c}
        2 J r^{2J-1}      
        \\
        L r^{2J-1}
        \\
        0
      \end{array}
    \right).
    \label{Psi1(v=0)}
  \end{align}
\item For the second set of mode functions, $\chi_2$ and $\eta_2$
  behave as
  \begin{align}
    \chi_2 (r\rightarrow\infty) &\, \simeq 
    a_1 r^{2J}
    -\frac{1}{2 m_\phi g^2 \bar{\phi}^2 r} f^{(\eta)}_J + \cdots,
    \label{chi2(v=0)}
    \\
    \eta_2 (r\rightarrow\infty) &\, = f^{(\eta)}_J,
  \end{align}
  with $a_1$ being a constant.  Here, we have used
  $f_J^{(\eta)}(r\rightarrow\infty)\propto r^{-2}$; this asymptotic
  behavior can be derived by using the fact that, in the left-hand
  side of Eq.\ \eqref{eq_feta}, the last term dominates when
  $r\rightarrow\infty$.  One can see that the leading order
  contributions to the top and the middle components of $\Psi_2$
  vanish.  For the calculation of the top and middle components, it is
  convenient to use Eqs.\ \eqref{Psitop'} and \eqref{Psimid'} to
  obtain
  \begin{align}
    \Psi_2 (r\rightarrow \infty) \simeq 
    a_1 \Psi_1
    + 
    \left( \begin{array}{c}
        \displaystyle{ -\frac{r}{L^2} f^{(\eta)}_J }
        \\[3mm]
        \displaystyle{ -\frac{2r}{L^3} f^{(\eta)}_J }
        \\[3mm]
        \displaystyle{ -\frac{1}{2 m_\phi g \bar{\phi} r} f^{(\eta)}_J }
      \end{array}
    \right).
    \label{Psi2(v=0)}
  \end{align}
  Notice that $\Psi_2^{\rm (bot)} (r\rightarrow \infty)\sim
  O(e^{m_\phi r}/r^{3/2})$.
\item For $\Psi_3$, we find
  \begin{align}
    \chi_3 (r\rightarrow\infty) &\, \simeq 
    b_1 r^{2J}
    - \frac{b_2}{2 m_\phi g^2 \bar{\phi}_C^2 r} f^{(\eta)}_J
    - \frac{1}{g^2 \bar{\phi}^2} r^{2J}
    + \cdots,
    \\
    \eta_3 (r\rightarrow\infty) &\, \simeq b_2 f^{(\eta)}_J
    + 2 J r^{2J}
    + \cdots,
    \\
    \zeta_3 (r\rightarrow\infty) &\, = r^{2J},
  \end{align}
  with $b_1$ and $b_2$ being constants.  Again, using Eqs.\
  \eqref{Psitop'} and \eqref{Psimid'},
  \begin{align}
    \Psi_3 (r\rightarrow \infty) \simeq 
    b_1 \Psi_1 + b_2 \Psi_2
    + 
    \left( \begin{array}{c}
        \displaystyle{ -\frac{(J +1)\xi-J}{4(J +1)} r^{2J+1} }
        \\[3mm]
        \displaystyle{ -\frac{2J[(J +1)\xi-(J +2)]}{4L(J +1)} r^{2J+1}}
        \\[3mm]
        O(\bar{\phi} r^{2J+2})
      \end{array}
    \right).
    \label{Psi3(v=0)}
  \end{align}
\end{enumerate}
Using Eqs.\ \eqref{Psi1(v=0)}, \eqref{Psi2(v=0)}, and
\eqref{Psi3(v=0)}, the determinant is given by
\begin{align}
  {\cal D}_J^{(S,L,\varphi)} (r\rightarrow\infty) = 
  \frac{ 4J^2 [ (J+1)\xi + J] r^{4J-1} f_J^{(\eta)}}{2 L^3 m_\phi g \bar{\phi}}.
  \label{D_l(v=0,r=inf)}
\end{align}

For the calculation of the functional determinants around the false
vacuum, we find the following solutions of Eq.\ \eqref{hatMl*Psi=0}:
\begin{align}
  \widehat\Psi_1 (r) &\, =
  \left( \begin{array}{c}
      2J r^{2J-1} \\ L r^{2J-1} \\ 0
    \end{array}
  \right),
  \\[2mm]
  \widehat\Psi_2 (r) &\, = 
  \left( \begin{array}{c}
      \displaystyle{ \frac{(J+1)\xi-J}{2L^2} r^{2J+1} }
      \\[3mm]
      \displaystyle{ \frac{(J+1)\xi-(J+2)}{4 L (J+1)} r^{2J+1} }
      \\[3mm]
      0
    \end{array}
  \right),
  \\[2mm]
  \widehat\Psi_3 (r) &\, = 
  \left( \begin{array}{c}
      0 \\ 0 \\ \widehat{f}_J^{(\sigma)} (r)
    \end{array}
  \right).
\end{align}
Here, the function $\widehat{f}^{(\sigma)}_J$ satisfies
\begin{align}
  (\Delta_J - m_\phi^2) \widehat{f}^{(\sigma)}_J = 0,
  \label{eq_fsighat}
\end{align}
with 
\begin{align}
  \widehat{f}^{(\sigma)}_J (r\rightarrow 0) \simeq 
  r^{2J}.
\end{align}
The explicit form of $\widehat{f}^{(\sigma)}_J$ is given by
\begin{align}
  \widehat{f}^{(\sigma)}_J (r) = 2^{2J + 1} \Gamma (2 J + 2)
  m_\phi^{-(2J +1)}
  \frac{I_{2J +1} (m_\phi r)}{r}.
\end{align}
Then, we find
\begin{align}
  \widehat{\cal D}_J^{(S,L,\varphi)} (r) = 
  - \frac{2J [ (J+1)\xi + J]}{L^3} r^{4J} \widehat{f}^{(\sigma)}_J (r).
  \label{hatD_l}
\end{align}

Using Eqs.\ \eqref{BC:D_J(S,L,NG)}, \eqref{D_l(v=0,r=inf)}, and
\eqref{hatD_l}, we obtain
\begin{align}
  \frac{\mbox{Det} {\cal M}_J^{(S,L,\varphi)}}
  {\mbox{Det} \widehat{\cal M}_J^{(S,L,\varphi)}} = 
  \frac{J\bar{\phi}_C}{m_\phi} 
  \frac{f_J^{(\eta)}(r_\infty)}
  {r_\infty \bar{\phi}(r_\infty) \widehat{f}_J^{(\sigma)}(r_\infty)}.
  \label{detratio(v=0)}
\end{align}
Importantly, the above result is $\xi$-independent.

We also give an alternative expression of the ratio of $\mbox{Det}
{\cal M}_J^{(S,L,\varphi)}$ to $\mbox{Det} \widehat{\cal
  M}_J^{(S,L,\varphi)}$, which is useful for the numerical
calculation.  Consider the limit of vanishing gauge coupling constant
$g$.  Even in such a limit, Eq.\ \eqref{hatD_l} is still valid if we
evaluate $f_J^{(\eta)}$ with $g=0$.  In addition, when $g=0$, the
fluctuation operator ${\cal M}_J^{(S,L,\varphi)}$ given in Eq.\
\eqref{M_J} is block-diagonal; the upper $2\times 2$ part becomes
independent of the bounce $\bar{\phi}$, leading to
\begin{align}
  \left[
    \frac{\mbox{Det} {\cal M}_J^{(S,L,\varphi)}}
    {\mbox{Det} \widehat{\cal M}_J^{(S,L,\varphi)}}
  \right]_{g=0} =
  \frac{\mbox{Det}[-\Delta_J+(\Delta_0 \bar{\phi})/\bar{\phi}]}
  {\mbox{Det}(-\Delta_J+m_\phi^2)}.
\end{align}
Thus, the following relation holds:
\begin{align}
  \frac{J\bar{\phi}_C}{m_\phi} 
  \frac{[f_J^{(\eta)}]_{g=0}(r_\infty)}
  {r_\infty \bar{\phi}(r_\infty) \widehat{f}_J^{(\sigma)}(r_\infty)}
  = 
  \frac{f^{(\sigma)}_J (r_\infty)}{\widehat{f}^{(\sigma)}_J (r_\infty)},
  \label{TwoMethods}
\end{align}
where the function $[f_J^{(\eta)}]_{g=0}$ obeys
\begin{align}
  \Delta_J [f_J^{(\eta)}]_{g=0}
  - \frac{2 \bar{\phi}'}{r^2 \bar{\phi}} \partial_r
  \left(
    r^2 [f_J^{(\eta)}]_{g=0}
  \right)
  = 0,
  \label{eq_feta(g=0)}
\end{align}
with 
\begin{align}
  [f_J^{(\eta)}]_{g=0} (r\rightarrow 0) \simeq r^{2J},
\end{align}
while $f^{(\sigma)}_J$ is the function which satisfies
\begin{align}
  \Delta_J f_J^{(\sigma)}
  - \frac{(\Delta_0 \bar{\phi})}{\bar{\phi}} f_J^{(\sigma)} = 0,
  \label{eq_fsig}
\end{align}
with
\begin{align}
  f_J^{(\sigma)} (r\rightarrow 0) \simeq r^{2J}.
\end{align}
Using Eq.\ \eqref{TwoMethods}, the ratio of the functional
determinants can be rewritten as
\begin{align}
  \frac{\mbox{Det} {\cal M}_J^{(S,L,\varphi)}}
  {\mbox{Det} \widehat{\cal M}_J^{(S,L,\varphi)}} = 
  \frac{f^{(\sigma)}_J (r_\infty)}{\widehat{f}^{(\sigma)}_J (r_\infty)}
  \frac{f_J^{(\eta)}(r_\infty)}{[f_J^{(\eta)}]_{g=0}(r_\infty)}.
  \label{detratioYS}
\end{align}
We note here that the evolution equations of the functions
$\widehat{f}^{(\sigma)}_J$ and $f^{(\sigma)}_J$, which are given in
Eqs.\ \eqref{eq_fsighat} and \eqref{eq_fsig}, respectively, are
asymptotically the same at $r\rightarrow\infty$; the same is true for
the evolution equations of the functions $[f_J^{(\eta)}]_{g=0}$ and
$f_J^{(\eta)}$.  Thus, the ratios
$f^{(\sigma)}_J/\widehat{f}^{(\sigma)}_J$ and
$f_J^{(\eta)}/[f_J^{(\eta)}]_{g=0}$ converge to constant values in the
limit of $r\rightarrow\infty$.  Numerically, the right-hand side of
Eq.\ \eqref{detratioYS} converges much faster than that of Eq.\
\eqref{detratio(v=0)} at $r\rightarrow\infty$.  Thus, Eq.\
\eqref{detratioYS} is useful for the numerical calculation.

\subsection{$v=0$: contribution of $J= 0$}

Now, we consider the functional determinants of ${\cal
  M}_{J=0}^{(S,\varphi)}$ and $\widehat{\cal M}_{J=0}^{(S,\varphi)}$
for the case of $v=0$.  When $v=0$, the $J=0$ mode requires special
treatment because all the classical solutions given in Eq.\
\eqref{ClassicalSolution}, parameterized by $\vartheta$, becomes the
bounce and contributes to the decay rate.  As a consequence, the
fluctuation operator ${\cal M}_{J=0}^{(S,\varphi)}$ has a zero mode
\cite{Kusenko:1996bv}.  Because of the zero mode, $\mbox{Det}{\cal
  M}_{J=0}^{(S,\varphi)}$ vanishes if one naively calculates the
functional determinant, resulting in a divergent behavior of the decay
rate of the false vacuum.  In the following, we discuss how to
calculate the $J=0$ contribution to obtain a meaningful result.

For the case of $J=0$, we need solutions of Eq.\
\eqref{M0*Psi=0}. From Eq.\ \eqref{eq_chi} (with $\eta=0$), the
following equations can be derived:
\begin{align}
  \Psi^{\rm (top)} = &\,
  - \xi \frac{1}{r^3} \int_0^r dr_1 r_1^3 \zeta (r_1),
  \label{Psitop'(J=0)}
  \\
  \Psi^{\rm (bot)} = &\,
  - \xi g \bar{\phi} 
  \left[
    c +
    \int^r dr_1 \frac{1}{r_1^3} 
    \int_0^{r_1} dr_2 r_2^3 \zeta (r_2)
  \right]
  + g \bar{\phi} \int^r dr_1
  \frac{1}{g^2 \bar{\phi}^2 (r_1)} \zeta' (r_1),
  \label{Psibot'(J=0)}
\end{align}
where $c$ is an arbitrary constant, and $\zeta'\equiv\partial_r\zeta$.
These equations are useful to derive the solutions of Eq.\
\eqref{M0*Psi=0}.  For $J=0$, the first solution of Eq.\
\eqref{M0*Psi=0} is obtained by taking $\chi_1(r)=1$ and
$\zeta_1(r)=0$:
\begin{align}
  \Psi_1 (r) = 
  \left( \begin{array}{c}
      0
      \\
      g \bar{\phi}
    \end{array}
  \right).
\end{align}
The second solution is obtained with $\chi_2(0)=0$ and $\zeta_2(r)=1$:
\begin{align}
  \Psi_2 (r) = 
  \left( \begin{array}{c}
      \displaystyle{ -\frac{1}{4} \xi r }
      \\[3mm]
      \displaystyle{ -\frac{1}{8} \xi r^2 g \bar{\phi} }
    \end{array}
  \right).
\end{align}
In addition, two independent solutions of $\widehat{\cal
  M}^{(S,\varphi)}_0\widehat{\Psi}=0$ are taken to be
\begin{align}
  \widehat{\Psi}_1 = &\,
  \left( \begin{array}{c}
      r
      \\
      0
    \end{array}
  \right),
  \\
  \widehat{\Psi}_2 = &\,
  \left(
    \begin{array}{c}
      0
      \\
      \widehat{f}_0^{(\sigma)}
    \end{array}
  \right).
\end{align}

One can see that $\Psi_1 (r\rightarrow\infty)\simeq 0$, and
$\mbox{Det}{\cal M}_{J=0}^{(S,\varphi)}=0$.  This is a consequence of
the zero-mode eigenfunction of ${\cal M}_{J=0}^{(S,\varphi)}$.
Indeed, the following function:
\begin{align}
  \Psi^{\rm (zero\mathchar"712Dmode)} 
  = {\cal N} {\cal Y}_{0,0,0}^{-1}
  \left( \begin{array}{c} 0 \\ \bar{\phi} \end{array} \right),
  \label{zeromode}
\end{align}
satisfies the conditions to be the zero-mode eigenfunction, i.e.,
${\cal M}^{(S,\varphi)}_0\Psi^{\rm (zero\mathchar"712Dmode)}=0$, and
$\Psi^{\rm (zero\mathchar"712Dmode)}(r\rightarrow\infty)=0$.  Here,
${\cal N}$ is the normalization factor, given by\footnote
{We adopt the normalization of the mode functions so that the path
  integral is defined as $\prod_n dc^{(n)}$, with $c^{(n)}$ being the
  expansion coefficient of the wave function with respect to the mode
  functions: $\Psi=\sum_n c^{(n)} \Psi^{(n)}$.  (In Eq.\
  \eqref{DPsi_zeromode}, $dc^{\rm (zero\mathchar"712Dmode)}$ is
  denoted as ${\cal D} \Psi^{\rm (zero\mathchar"712Dmode)}$.)  Some of
  the previous studies use different definition of the path integral
  as $\prod_n (dc^{(n)}/\sqrt{2\pi})$, with which the right-hand side
  of Eq.\ \eqref{normalization} should be $1$. }
\begin{align}
  {\cal N}^2 \int d^4 r \bar{\phi}^2 = 2 \pi.
  \label{normalization}
\end{align}

The zero mode stems from the global $U(1)$ symmetry which is preserved
in the false vacuum; such a $U(1)$ symmetry relates the bounce
configurations parameterized by $\vartheta$ (see Eq.\
\eqref{bounce_v=0}).  The zero mode given in Eq.\ \eqref{zeromode} is
nothing but the mode generated by the global $U(1)$ transformation of
the bounce.  Thus, the path integral of the zero mode should be
understood as the integration over the bounce configurations related
by the $U(1)$ transformation.  Based on this consideration, we can
perform the following replacements \cite{Kusenko:1996bv}:
\begin{align}
  \int {\cal D} \Psi^{\rm (zero\mathchar"712Dmode)} \rightarrow
  \frac{1}{\cal N} \int_0^{2\pi} d\vartheta,
  \label{DPsi_zeromode}
\end{align}
where $\int {\cal D} \Psi^{\rm (zero\mathchar"712Dmode)}$ denotes the
path integral of the zero mode, and, using the fact that the
fluctuation operators do not depend on $\vartheta$,
\begin{align}
  \left[
    \frac{\mbox{Det}{\cal M}^{(S,\varphi)}_0}
    {\mbox{Det}\widehat{\cal M}^{(S,\varphi)}_0} 
  \right]^{-1/2}
  \rightarrow
  \frac{2\pi}{\cal N}
  \left[
    \frac{\mbox{Det}'{\cal M}^{(S,\varphi)}_0}
    {\mbox{Det}\widehat{\cal M}^{(S,\varphi)}_0}
  \right]^{-1/2},
  \label{ZeromodeIntegration}
\end{align}
where $\mbox{Det}'$ implies that the zero eigenvalue is omitted from
the functional determinant.

The zero eigenvalue can be omitted with the use of the following
modified fluctuation operator:
\begin{align*}
  {\cal M}^{(S,\varphi)}_0  + \mbox{diag} (\nu,\nu),
\end{align*}
where $\nu$ is a (small) constant.  Each eigenfunction of ${\cal
  M}^{(S,\varphi)}_0$ is also an eigenfunction of the above modified
fluctuation operator; the eigenvalue increases by $\nu$.  Especially,
$\Psi^{\rm (zero\mathchar"712Dmode)}$ given in Eq.\ \eqref{zeromode}
is an eigenfunction of the above modified fluctuation operator with
the eigenvalue of $\nu$.  Thus, we eliminate the zero eigenvalue from
$\mbox{Det}{\cal M}^{(S,\varphi)}_0$ as
\begin{align}
  \frac{\mbox{Det}'{\cal M}^{(S,\varphi)}_0}
  {\mbox{Det}\widehat{\cal M}^{(S,\varphi)}_0} 
  =\lim_{\nu\rightarrow 0} \frac{1}{\nu}
  \frac{\mbox{Det} [{\cal M}^{(S,\varphi)}_0 + \mbox{diag} (\nu,\nu)]}
  {\mbox{Det}\widehat{\cal M}^{(S,\varphi)}_0}.
\end{align}

For the calculation of the functional determinant of the modified
fluctuation operator, we solve the following equation:
\begin{align}
  \left[ {\cal M}^{(S,\varphi)}_0 + \mbox{diag} (\nu,\nu) \right]
  \Psi^{(\nu)} = 0,
\end{align}
with the condition $\lim_{\nu\rightarrow 0}\Psi^{(\nu)}=\Psi_1$.
Defining
\begin{align}
  {\cal D}'^{(S,\varphi)}_0 (r) = 
  \lim_{\nu\rightarrow 0}
  \frac{\mbox{det}(\Psi^{(\nu)} (r)~\Psi_2 (r))}{\nu},
\end{align}
we obtain
\begin{align}
  \frac{\mbox{Det}'{\cal M}^{(S,\varphi)}_0}
  {\mbox{Det}\widehat{\cal M}^{(S,\varphi)}_0}
  = 
  \left[
    \frac{{\cal D}_0^{(S,\varphi)} (r\rightarrow 0)}
    {\widehat{\cal D}_0^{(S,\varphi)} (r\rightarrow 0)}
  \right]^{-1}
  \frac{{\cal D}'^{(S,\varphi)}_0 (r_\infty)}
  {\widehat{\cal D}_0^{(S,\varphi)} (r_\infty)}.
\end{align}

For the calculation of $\Psi^{(\nu)}$ up to $O(\nu)$, we expand
$\Psi^{(\nu)}$ as
\begin{align}
  \Psi^{(\nu)} = \Psi_1 + \nu \check{\Psi} + O(\nu^2),
\end{align}
with which
\begin{align}
  {\cal D}'^{(S,\varphi)}_0(r_\infty) = 
  \mbox{det}(\check{\Psi} (r_\infty)~\Psi_2 (r_\infty)).
\end{align}
Here, $\check{\Psi}$ should satisfy
\begin{align}
  {\cal M}^{(S,\varphi)}_0 \check{\Psi} =
  - \left( \begin{array}{c}
      0 \\ g\bar{\phi}
    \end{array} \right).
\end{align}
The solution of the above equation is given in the following form:
\begin{align}
  \check{\Psi} = 
  \left( \begin{array}{c}
      \check{\Psi}^{\rm (top)} \\
      \check{\Psi}^{\rm (bot)}
    \end{array} 
  \right) 
  \equiv
  \left( \begin{array}{c}
      \partial_r \check{\chi}
      \\
      g \bar{\phi} \check{\chi}
    \end{array} 
  \right) 
  + 
  \left( \begin{array}{c}
      \displaystyle{
        -2 \frac{\bar{\phi}'}{g^2 \bar{\phi}^3} \check{\zeta}
      }
      \\[3mm] 
      \displaystyle{ \frac{1}{g \bar{\phi}} \check{\zeta} }
    \end{array} 
  \right),
\end{align}
with the functions $\check{\chi}$ and $\check{\zeta}$ obeying
\begin{align}
  &
  \Delta_0 \check{\chi} = 
  \frac{2}{r^3} \partial_r 
  \left( \frac{r^3 \bar{\phi}'}{g^2 \bar{\phi}^3} \check{\zeta} \right)
  - \xi \check{\zeta},
  \\[2mm] &
  \Delta_0 \check{\zeta} = g^2 \bar{\phi}^2.
  \label{eq_tildezeta}
\end{align}
Notice that, from Eq.\ \eqref{eq_tildezeta}, $\check{\zeta}$ is given by
\begin{align}
  \check{\zeta} = 
  \int_0^r dr_1 r_1^{-3} \int_0^{r_1} dr_2 r_2^3 g^2 \bar{\phi}^2 (r_2),
\end{align}
and hence $\check{\zeta}(r\rightarrow\infty)$ is a constant.  Notably,
$\check{\Psi}^{\rm (top)}$ and $\check{\Psi}^{\rm (bot)}$ satisfy
similar equations as Eqs.\ \eqref{Psitop'(J=0)} and
\eqref{Psibot'(J=0)}, respectively:
\begin{align}
  \check{\Psi}^{\rm (top)} (r) = &\,
  - \xi \frac{1}{r^3} \int_0^r dr_1 r_1^3 \check{\zeta} (r_1),
  \label{tildePsitop'(J=0)}
  \\
  \check{\Psi}^{\rm (bot)} (r) = &\,
  - \xi g \bar{\phi} \int^r dr_1 \frac{1}{r_1^3} 
  \int_0^{r_1} dr_2 r_2^3 \check{\zeta} (r_2)
  + g \bar{\phi} \int^r dr_1
  \frac{1}{g^2 \bar{\phi}^2 (r_1)} \check{\zeta}' (r_1).
  \label{tildePsibot'(J=0)}
\end{align}
We are interested in their behaviors at $r\rightarrow\infty$; in such
a limit, (i) $\check{\Psi}^{\rm (top)}$ is proportional to $r$ because
$\check{\zeta}(r\rightarrow\infty)$ is a constant, and (ii) the
asymptotic behavior of $\check{\Psi}^{\rm (bot)}$ is obtained from the
fact that the first term of the right-hand side of Eq.\
\eqref{tildePsibot'(J=0)} vanishes when $r\rightarrow\infty$ and that
$\check{\zeta}'(r\rightarrow\infty)\simeq g^2/\pi {\cal N}^2r^3$.
Remembering that $\bar{\phi}(r\rightarrow\infty)$ is approximately
proportional to $e^{-m_\phi r}/r^{3/2}$, the asymptotic behavior of
$\check{\Psi}$ is found to be
\begin{align}
  \check{\Psi} (r\rightarrow\infty) \simeq
  \left( \begin{array}{c}
      \displaystyle{ -\frac{1}{4} \xi r \check{\zeta}}
      \\[3mm] 
      \displaystyle{
        \frac{g}{2\pi {\cal N}^2 m_\phi r^3 \bar{\phi}}
      }
    \end{array} 
  \right).
\end{align}

Consequently, we obtain
\begin{align}
  \left[
    \frac{\mbox{Det}{\cal M}^{(S,\varphi)}_0}
    {\mbox{Det}\widehat{\cal M}^{(S,\varphi)}_0} 
  \right]^{-1/2}
  \rightarrow
  2 \pi
  \left[
    \frac{1}
    {2\pi m_\phi \bar{\phi}_C r_\infty^3 \bar{\phi} (r_\infty)
      \widehat{f}^{(\sigma)}_0 (r_\infty)}
  \right]^{-1/2},
\end{align}
where we have used $f^{(\sigma)}_0(0)=1$.  Notice that
$\widehat{f}^{(\sigma)}_0 (r\rightarrow\infty)\propto e^{m_\phi
  r}/r^{3/2}$, and that the above quantity is finite.

\subsection{$v=0$: final result}

The contributions of the $S$, $L$, and NG modes are
\begin{align}
  {\cal A}^{(S,L,\varphi)}
  = 
  2 \pi
  \left[
    2\pi m_\phi \bar{\phi}_C r_\infty^3 \bar{\phi} (r_\infty)
    \widehat{f}^{(\sigma)}_0 (r_\infty)
  \right]^{1/2}
  \prod_{J\geq 1/2}
  \left[
    \frac{J\bar{\phi}_C}{m_\phi} 
    \frac{f_J^{(\eta)}(r_\infty)}
    {r_\infty \bar{\phi}(r_\infty) \widehat{f}_J^{(\sigma)}(r_\infty)}
  \right]^{-(2J +1)^2/2},
  \label{A^SLphi(final):v=0}
\end{align}
while the other contributions are
\begin{align}
  {\cal A}^{(T)} =&\,
  \prod_{J=1/2}^{\infty} \left[
    \frac{f^{(T)}_J (r_\infty)}{\widehat{f}^{(T)}_J (r_\infty)}
  \right]^{-(2J +1)^2},
  \label{A^T(final):v=0}
  \\
  {\cal A}^{(\bar{c},c)} =&\,
  1.
\end{align}
For the case of $v=0$, $\widehat{f}^{(\eta)}_J (r_\infty)\propto
r^{-2}$ and $\widehat{f}^{(T)}_J (r_\infty)\propto r^{2J}$.  We
emphasize that the final result is $\xi$-independent.

Before closing this section, we comment on the calculation based on
the $R_\xi$-like gauge fixing function, ${\cal F}^{(R_\xi)}
= \partial_\mu A_\mu - 2 \xi g (\mbox{Re}\Phi)(\mbox{Im}\Phi)$.  As we
have mentioned, in the case of $v=0$, there exists a class of bounce
configuration which depends on the function $\Theta (r)$ obeying Eq.\
\eqref{eq_Theta}; the function $\Theta$ is determined by its value at
the origin, $\Theta (0)$.  One technical difficulty is that the bounce
configurations, as well as the fluctuation operators around the
bounce, depend on $\Theta (0)$. If we adopt ${\cal F}^{(R_\xi)}$, we
need to calculate the functional determinants as functions of $\Theta
(0)$, and somehow integrate over $\Theta (0)$.  Such an analysis is
beyond the scope of this paper, because we have shown that the final
result can be obtained with the use of the gauge fixing function
${\cal F} = \partial_\mu A_\mu$.  Just for a comparison, we have
calculated the functional determinants around the bounce configuration
with $\Theta (0)=0$ (which results in $\Theta=0$), adopting ${\cal
  F}^{(R_\xi)}$.  Based on the calculation with angular-momentum
decomposition, we have checked that the contributions from the modes
with $J\neq 0$ agree with the results of the present calculation.
However, the contribution of the $J=0$ mode is hardly compared with
our present result because the measure for the integration over
$\Theta (0)$ is unknown.  Notice that a hasty substitution of
$\mbox{Det}'[{\cal
  M}^{(S,\varphi)}_0]_{R_\xi}/\mbox{Det}[\widehat{\cal
  M}^{(S,\varphi)}_0]_{R_\xi}$ for $\Theta=0$ into Eq.\
\eqref{ZeromodeIntegration} will give a gauge dependent result, where
$[{\cal M}^{(S,\varphi)}_0]_{R_\xi}$ and $[\widehat{\cal
  M}^{(S,\varphi)}_0]_{R_\xi}$ are fluctuation operators based on the
$R_\xi$-like gauge.

\section{Renormalization}
\label{sec:renormalization}
\setcounter{equation}{0}

So far, we have calculated the functional determinants by integrating
out the field fluctuations around the bounce configuration and also
around the false vacuum.  Because these quantities are divergent
\cite{Isidori:2001bm}, the renormalization is necessary to make the decay
rate finite.  In this section, we outline how to perform
the renormalization.  As in the previous sections, we pay particular
attention to the effects of the gauge bosons and NG boson.

First, for notational simplicity, we introduce
\begin{align}
  \delta \bar{\phi}^2 \equiv \bar{\phi}^2 - v^2,
\end{align}
and 
\begin{align}
  \delta \Omega \equiv \Omega - \widehat{\Omega},
\end{align}
with
\begin{align}
  \Omega \equiv \frac{(\partial^2 \bar{\phi})}{\bar{\phi}}
  + g^2 \bar{\phi}^2.
\end{align}
Here, $\widehat{\Omega}$ is the value of $\Omega$ around the false
vacuum (and hence is a constant):
\begin{align}
  \widehat{\Omega} = 
  \left\{
    \begin{array}{ll}
      g^2 v^2 & :~ v\neq 0
      \\
      m_\phi^2 & :~ v = 0
    \end{array}
  \right. .
\end{align}
The calculation of the functional determinant of ${\cal
  M}^{(A_\mu,\varphi)}$ can be performed by treating
$\delta\bar{\phi}^2$ and $\delta\Omega$ as perturbations; ${\cal
  M}^{(A_\mu,\varphi)}$ is given by the sum of the terms with
different numbers of the insertions of $\delta\bar{\phi}^2$ and
$\delta\Omega$.

Because we are interested in renormalizable theories, all the
divergences are related to operators with mass dimension $4$ or
smaller.  In the present model, such divergences show up at finite
orders of the gauge or quartic scalar couplings at the one-loop level.
In other words, the divergences are with limited numbers of the
insertions of $\delta\bar{\phi}^2$ and $\delta\Omega$.  The procedure
to obtain renormalized decay rate is to calculate the functional
determinant without the divergent part, which will be defined below,
by the method adopted in the previous sections.  The divergent part is
separately calculated with the dimensional regularization using
ordinary Feynman rules, and is made finite with the
$\overline{\mbox{MS}}$ subtraction.

The divergent part can be obtained by expanding the functional
determinant with respect to $\delta\bar{\phi}^2$ and $\delta\Omega$,
and keeping the terms corresponding to operators with mass dimensions
less than or equal to $4$.  Importantly, the divergent part should be
properly subtracted from the functional determinant for each $J$.  It
may be performed with the fluctuation operators given in Eqs.\
\eqref{M_J} and \eqref{M_J=0}, which are obtained from the gauge
fixing function of our choice.  However, the calculation can be made
easier if we use the fluctuation operators obtained from the
$R_\xi$-like gauge fixing function; it is allowed because we have
confirmed that the results of the calculations based on the two
different gauge fixing functions agree with each other at least for
$J\neq 0$ \cite{Endo:2017gal}.  Hereafter, we use these facts to
simplify our calculation.

With a straightforward calculation, the divergent part of
$\ln[{\mbox{Det} {\cal M}^{(A_\mu,\varphi)}}/{\mbox{Det} \widehat{\cal
    M}^{(A_\mu,\varphi)}}]^{-1/2}$ is obtained as
\begin{align}
  \delta {\cal S}^{(A_\mu,\varphi)}_{\rm div} \equiv &\, 
  \mbox{Tr} 
  \left[ 
    g^2 \delta\bar{\phi}^2 
    \frac{1}{-\partial^2+g^2 v^2}
  \right]
  - \frac{1}{2} \mbox{Tr} 
  \left[
    g^2 \delta\bar{\phi}^2 
    \frac{1}{-\partial^2+g^2 v^2}
    g^2 \delta\bar{\phi}^2 
    \frac{1}{-\partial^2+g^2 v^2}
  \right]
  \nonumber \\ &\, 
  + \frac{1}{2} 
  \mbox{Tr} 
  \left[ 
    \delta\Omega
    \frac{1}{-\partial^2+\widehat{\Omega}}
  \right]
  -\frac{1}{4} \mbox{Tr} 
  \left[
    \delta\Omega
    \frac{1}{-\partial^2+\widehat{\Omega}}
    \delta\Omega
    \frac{1}{-\partial^2+\widehat{\Omega}}
  \right]
  \nonumber \\ &\, -2
  \mbox{Tr} 
  \left[ 
    (g\partial_\mu \bar{\phi}) \frac{1}{-\partial^2+g^2 v^2}
    (g\partial_\mu \bar{\phi}) \frac{1}{-\partial^2+\widehat{\Omega}}
  \right],
\end{align}
which means
\begin{align}
  \ln \left[
    \frac{\mbox{Det} {\cal M}^{(A_\mu,\varphi)}}
    {\mbox{Det} \widehat{\cal M}^{(A_\mu,\varphi)}}
  \right]^{-1/2}
  + \delta{\cal S}^{(A_\mu,\varphi)}_{\rm div}
  = (\mbox{finite}).
\end{align}
As we see below, $\delta{\cal S}^{(A_\mu,\varphi)}_{\rm div}$ can be
used to subtract the divergences with a relevant renormalization scheme
(like the $\overline{\mbox{MS}}$ scheme).  Notice that the divergent
part of the functional determinant of our interest is
$\xi$-independent, and hence the decay rate of the false vacuum is
gauge invariant even after the renormalization.

Hereafter, we calculate $\delta{\cal S}^{(A_\mu,\varphi)}_{\rm div}$
with two different procedures.  One is a decomposition with respect
to the angular momentum, which is based on the following equality:
\begin{align}
  \delta {\cal S}^{(A_\mu,\varphi)}_{\rm div} =
  \delta {\cal S}^{(S,L,\varphi)}_{\rm div} +
  \delta {\cal S}^{(T)}_{\rm div},
\end{align}
where
\begin{align}
  \delta {\cal S}^{(S,L,\varphi)}_{\rm div} \equiv &\,
  \frac{1}{2} \left[ \ln
    \frac{\mbox{Det}(\widehat{\cal M}_{R_{\xi =1}}^{(A_\mu,\varphi)}
      + \delta {\cal M})}
    {\mbox{Det}\widehat{\cal M}_{R_{\xi =1}}^{(A_\mu,\varphi)}}
  \right]_{(\delta {\cal M})^2}
  - 2 \left[
    \ln 
    \frac{\mbox{Det}
      ( -\partial^2 + g^2 v^2 + g^2 \delta\bar{\phi}^2 )
    }
    {\mbox{Det} (-\partial^2 + g^2 v^2)}
  \right]_{(\delta\bar{\phi}^2)^2},
  \label{Sdiv_SLverphi}
  \\
  \delta {\cal S}^{(T)}_{\rm div}\equiv &\,
  \left[
    \ln 
    \frac{\mbox{Det}
      ( -\partial^2 + g^2 v^2 + g^2 \delta\bar{\phi}^2 )
    }
    {\mbox{Det} (-\partial^2 + g^2 v^2)}
  \right]_{(\delta\bar{\phi}^2)^2},
  \label{Sdiv_T}
\end{align}
with
\begin{align}
  \widehat{\cal M}_{R_{\xi =1}}^{(A_\mu,\varphi)} \equiv &\,
  \left(
    \begin{array}{cc}
      (- \partial^2 + g^2 v^2) \delta_{\mu\nu}
      & 0
      \\
      0
      & - \partial^2 + \widehat{\Omega}
    \end{array}
  \right),
  \\
  \delta {\cal M} \equiv &\,
  \left(
    \begin{array}{cc}
      g^2 \delta\bar{\phi}^2  \delta_{\mu\nu}
      & 2 g (\partial_\nu \bar{\phi})
      \\
      2 g (\partial_\mu \bar{\phi}) 
      & \delta\Omega
    \end{array}
  \right).
\end{align}
Here, $[\cdots]_{{\cal P}^N}$ indicates that the quantity in the
square bracket is evaluated up to $O({\cal P}^N)$.  We note that
$\widehat{\cal M}_{R_{\xi =1}}^{(A_\mu,\varphi)}$ is the fluctuation
operator of $A_\mu$ and $\varphi$ in the $R_\xi$-like gauge with
$\xi=1$ around the false vacuum.  In addition, $\widehat{\cal
  M}_{R_{\xi =1}}^{(A_\mu,\varphi)}+\delta{\cal M}$ is the one around
the bounce configuration with $\Theta(r)=0$.  Because the results
based on our choice of the gauge fixing and the $R_\xi$-like gauge
fixing give the same result, Eqs.\ \eqref{Sdiv_SLverphi} and
\eqref{Sdiv_T} properly take account of the divergent part for each
$J$.  With Eq.\ \eqref{Gelfand-Yaglom}, the right-hand sides of Eqs.\
\eqref{Sdiv_SLverphi} and \eqref{Sdiv_T} can be evaluated.  The result
is given as the sum of the contribution from each angular momentum.
We denote
\begin{align}
  \delta{\cal S}^{(S,L,\varphi)}_{\rm div}
  \equiv &\,
  \sum_{J=0}^\infty s_J^{(S,L,\varphi)},
  \label{Sdiv_SLverphi(J)}
  \\
  \delta{\cal S}^{(T)}_{\rm div}
  \equiv &\,
  \sum_{J=0}^\infty s_J^{(T)}.
  \label{Sdiv_T(J)}
\end{align}
Notice that, comparing Eq.\ \eqref{Sdiv_T} with Eq.\ \eqref{M^T}, we
can see that $\ln {\cal A}^{(T)}+\delta{\cal S}^{(T)}_{\rm div}$ is
finite.  Thus, $\ln {\cal A}^{(S,L,\varphi)}+\delta {\cal
  S}^{(S,L,\varphi)}_{\rm div}$ also is.  A prescription for the
calculation of the counter terms for each angular momentum, i.e.,
$s_J^{(S,L,\varphi)}$ and $s_J^{(T)}$, is given in Appendix
\ref{app:finiteorder} (see Eqs.\ \eqref{Sdiv_SLverphiWithL} and
\eqref{Sdiv_TWithL}).
  
The quantity ${\cal S}^{(A_\mu,\varphi)}_{\rm div}$ is also calculated
with ordinary Feynman rules.  The result is divergent; using the
dimensional regularization based on $D$-dimensional theory, ${\cal
  S}^{(A_\mu,\varphi)}_{\rm div}$ constrains a term proportional to
$\bar{\epsilon}^{-1}\equiv\frac{2}{4-D}-\gamma+\ln 4\pi$ (with
$\gamma$ here being the Eular's constant).  Such a term is exactly
cancelled out by the counter term in the $\overline{\rm MS}$-scheme.
We define $\delta {\cal S}^{(A_\mu,\varphi)}_{\overline{\rm MS}}$ from
$\delta {\cal S}^{(A_\mu,\varphi)}_{\rm div}$ via the $\overline{\rm
  MS}$-subtraction:
\begin{align}
  \delta {\cal S}^{(A_\mu,\varphi)}_{\overline{\rm MS}} \equiv
  \left.
    \delta {\cal S}^{(A_\mu,\varphi)}_{\rm div}
  \right|_{\overline{\rm MS}{\rm \mathchar"712Dsubtraction}},
\end{align}
which is finite.  With the bounce solution $\bar{\phi}$, the explicit
expression of $\delta{\cal S}^{(A_\mu,\varphi)}_{\overline{\rm MS}}$
is given by
\begin{align}
  \delta {\cal S}^{(A_\mu,\varphi)}_{\overline{\rm MS}} = &\,
  \left[ g^2 \delta\bar{\phi}^2 \right]_{\rm FT} (0) 
  {\cal I}_1 (g^2 v^2)
  \nonumber \\ &\,
  -\frac{1}{2} \int \frac{d^4 k}{(2\pi)^4}
  \left[ g^2 \delta\bar{\phi}^2 \right]_{\rm FT} (-k) 
  \left[ g^2 \delta\bar{\phi}^2 \right]_{\rm FT} (k) 
  {\cal I}_2 (k^2; g^2v^2, g^2v^2)
  \nonumber \\ &\,
  +\frac{1}{2} 
    \left[ \delta\Omega \right]_{\rm FT} (0)
  {\cal I}_1(\widehat{\Omega})
  \nonumber \\ &\,
  -\frac{1}{4} \int \frac{d^4 k}{(2\pi)^4}
  \left[ {\delta\Omega} \right]_{\rm FT} (-k) 
  \left[ {\delta\Omega} \right]_{\rm FT} (k) 
  {\cal I}_2 (k^2; \widehat{\Omega}, \widehat{\Omega})
  \nonumber \\ &\,
  -2 \int \frac{d^4 k}{(2\pi)^2}
  [g \bar{\phi}]_{\rm FT} (-k) [g \bar{\phi}]_{\rm FT} (k) 
  k^2 {\cal I}_2 (k^2; g^2 v^2, \widehat{\Omega}),
\end{align}
where $[F]_{\rm FT}(k)$ is the Fourier transformation of the function
$F(x)$:
\begin{align}
  [F]_{\rm FT} (k) \equiv \int d^4 x e^{ikx} F (x),
\end{align}
and the loop functions ${\cal I}_1$ and ${\cal I}_2$ are given by
\begin{align}
  16\pi^2 {\cal I}_1 (m^2) \equiv &\,
  m^2 \left( \ln \frac{m^2}{\mu^2} - 1 \right),
  \\
  16\pi^2 {\cal I}_2 (k^2; m_1^2,m_2^2) \equiv &\,
  -\frac{1}{2}  \ln \frac{m_1^2}{\mu^2}
  -\frac{1}{2}  \ln \frac{m_2^2}{\mu^2}
  + 2
  - \frac{m_1^2 - m_2^2}{2k^2}
  \ln \frac{m_2^2}{m_1^2}
  \nonumber \\ &\,
  - \frac{\beta (k^2; m_1^2,m_2^2)}{2}
  \ln
  \frac{k^2+m_1^2+m_2^2+k^2 \beta (k^2; m_1^2,m_2^2)}
  {k^2+m_1^2+m_2^2 - k^2 \beta (k^2; m_1^2,m_2^2)},
\end{align}
with 
\begin{align}
  \beta (k^2; m_1^2,m_2^2) \equiv
  \frac{\sqrt{(k^2)^2+2k^2(m_1^2+m_2^2)+(m_1^2-m_2^2)^2}}{k^2},
\end{align}
and $\mu$ being the renormalization scale.

Because the bounce configuration is $O(4)$ symmetric, we can simplify
the expression of $\delta {\cal S}^{(A_\mu,\varphi)}_{\overline{\rm
    MS}}$.  Redefining $k=\sqrt{k_\mu k_\mu}$, the Fourier
transformations of spherically symmetric functions are given by
\begin{align}
  [F]_{\rm FT} (k) = 4 \pi^2 \int_0^\infty dr r^3 
  \frac{J_1 (kr)}{kr} F (r).
\end{align}
Then,
\begin{align}
  \delta {\cal S}^{(A_\mu,\varphi)}_{\overline{\rm MS}} = &\,
  \left[ g^2 \delta\bar{\phi}^2 \right]_{\rm FT} (0) 
  {\cal I}_1 (g^2 v^2)
  -\frac{1}{2} \int_0^\infty \frac{k^3 dk}{8\pi^2}
  \left[ g^2 \delta\bar{\phi}^2 \right]_{\rm FT}^2 (k) 
  {\cal I}_2 (k^2; g^2v^2, g^2v^2)
  \nonumber \\ &\,
  +\frac{1}{2} 
    \left[ {\delta\Omega} \right]_{\rm FT} (0)
  {\cal I}_1(\widehat{\Omega})
  -\frac{1}{4} \int_0^\infty \frac{k^3 dk}{8\pi^2}
  \left[ {\delta\Omega} \right]_{\rm FT}^2 (k) 
  {\cal I}_2 (k^2; \widehat{\Omega}, \widehat{\Omega})
  \nonumber \\ &\,
  -2 \int_0^\infty \frac{k^3 dk}{8\pi^2}
  [g \bar{\phi}]_{\rm FT}^2 (k)
  k^2 {\cal I}_2 (k^2; g^2 v^2, \widehat{\Omega}).
\end{align}

Based on the above argument, the functional determinant is
renormalized as follows:
\begin{align*}
  \left[
    \mbox{Det} {\cal M}^{(A_\mu,\varphi)}
  \right]^{-1/2}
  \rightarrow
  e^{-\delta {\cal S}^{(A_\mu,\varphi)}_{\overline{\rm MS}}}
  e^{\delta {\cal S}^{(A_\mu,\varphi)}_{\rm div}}
  \left[
    \mbox{Det} {\cal M}^{(A_\mu,\varphi)}
  \right]^{-1/2}.
\end{align*}
The decay rate of the false vacuum is evaluated as $\gamma = {\cal A}
e^{-{\cal B}}$, with
\begin{align}
  {\cal A} = 
  \frac{{\cal B}^2}{4\pi^2} {\cal A}'^{(h)}_{\rm R}
  {\cal A}^{(S, L, \varphi)}_{\rm R}
  {\cal A}^{(T)}_{\rm R}
  {\cal A}^{(\bar{c},c)}_{\rm R}
  e^{-\delta {\cal S}^{\rm (tot)}_{\overline{\rm MS}}}.
  \label{Atot(final)}
\end{align}
Here, 
\begin{align}
  \delta {\cal S}^{\rm (tot)}_{\overline{\rm MS}}
  = \delta {\cal S}^{(A_\mu,\varphi)}_{\overline{\rm MS}} + \cdots,
\end{align}
where ``$\cdots$'' indicates the contributions from the fields other
than $A_\mu$ and $\varphi$ (i.e., the Higgs mode, for example).  In
addition, the subscript ``R'' is for ``renormalized'' objects after
subtracting the divergences.  For the $S$, $L$, and NG modes,
\begin{align}
  {\cal A}^{(S,L,\varphi)}_{\rm R} = 
  e^{s^{(S,L,\varphi)}_0}
  \left[
    \frac{\mbox{Det} {\cal M}_0^{(S,\varphi)}}
    {\mbox{Det} \widehat{\cal M}_0^{(S,\varphi)}}
  \right]^{-1/2}
  \prod_{J=1/2}^{\infty} 
  e^{s^{(S,L,\varphi)}_J}
  \left[
    \frac{\mbox{Det} {\cal M}_J^{(S,L,\varphi)}}
    {\mbox{Det} \widehat{\cal M}_J^{(S,L,\varphi)}}
  \right]^{-(2J +1)^2/2},
\end{align}
while the contribution of the transverse mode is
\begin{align}
  {\cal A}^{(T)}_{\rm R} =
  e^{s_0^{(T)}}
  \prod_{J=1/2}^{\infty} 
  e^{s^{(T)}_J}
  \left[
    \frac{\mbox{Det} {\cal M}_J^{(T)}}
    {\mbox{Det} \widehat{\cal M}_J^{(T)}}
  \right]^{-(2J +1)^2}.
\end{align}
The expressions for the functional determinants for the case of $v\neq
0$ ($v=0$) are given in Eqs.\ \eqref{A^SLphi(final)} and
\eqref{A^T(final)} (Eqs.\ \eqref{A^SLphi(final):v=0} and
\eqref{A^T(final):v=0}).  Furthermore, the ghost contribution is
\begin{align}
  {\cal A}^{(\bar{c},c)}_{\rm R} = 1.
  \label{Ac(final)}
\end{align}
Obviously, the decay rate is gauge invariant even after
the renormalization.

\section{Conclusions and Discussion}
\label{sec:conclusion}
\setcounter{equation}{0}

In this paper, we have studied the false vacuum decay in gauge theory,
paying particular attention to the gauge invariance of the decay rate
of the false vacuum.  Using the model with $U(1)$ gauge symmetry, for
which the scalar field responsible for the metastability of the false
vacuum has non-vanishing charge, we have shown that the decay rate of
the false vacuum is indeed gauge invariant at least at the one-loop
level.  We have adopted the gauge fixing function of the form ${\cal
  F}=\partial_\mu A_\mu$.  We emphasize that such a choice of gauge
fixing function is advantageous when the gauge symmetry is preserved
in the false vacuum.  This is because the zero-mode fluctuation in
association with the internal symmetry can be successfully integrated
out.  Such an integration was hardly performed with the $R_\xi$-like
gauge fixing function.  We have also discussed a procedure to perform
the renormalization to remove divergences, and have shown that the decay
rate of the false vacuum is gauge independent even after
the renormalization.

Our main results are summarized at the end of Section
\ref{sec:renormalization} (see Eqs.\ \eqref{Atot(final)} $-$
\eqref{Ac(final)}).  The decay rate of the false vacuum is related to
the asymptotic values of the solutions of the second-order
differential equations which are gauge independent (i.e.,
$f^{(\eta)}$, $f^{(T)}$, and so on); for a given scalar potential with
a false vacuum, the second-order differential equations can be solved
numerically once the bounce configuration is determined.  Our results
simplify the numerical calculation of the decay rate because we only have
to study evolution of the gauge-invariant functions which do not mix
with the other functions.  In a brute-force calculation, on the
contrary, one should solve simultaneous differential equations
containing unphysical modes, which makes the numerical calculation
unstable.

Our results would have various phenomenological applications because
false vacuum decay and phase transition are important subjects in
particle physics and cosmology.  For example, with the measurements of
the Higgs mass at the LHC experiment \cite{Aad:2015zhl} as well as
those of top mass \cite{ATLAS:2014wva}, it has been realized that the
standard-model Higgs potential becomes unstable when the Higgs
amplitude becomes extremely large \cite{Isidori:2001bm,
  Degrassi:2012ry, Alekhin:2012py, Espinosa:2015qea,
  Plascencia:2015pga, Lalak:2016zlv, Espinosa:2016nld}.\footnote
{In the case of the Higgs potential of the standard model, which is
  dominated by the quartic term when calculating the decay rate of the
  false vacuum, the asymptotic behavior of the bounce at
  $r\rightarrow\infty$ is different from that adopted in the present
  analysis.  Such a case will be studied elsewhere.}
In various models of physics beyond the standard model, the metastable
vacua also show up.  Supersymmetry is one of the important examples
because there may exist color and/or charge breaking minima of the
scalar potential \cite{Gunion:1987qv, Casas:1995pd, Kusenko:1996jn,
  Hisano:2010re, Camargo-Molina:2013sta, Chowdhury:2013dka,
  Blinov:2013fta, Camargo-Molina:2014pwa, Endo:2015oia, Endo:2015ixx}.

\vspace{3mm}
\noindent {\it Acknowledgements}: The authors thank P. Cox for his
careful reading of the manuscript.  This work was supported by the
Grant-in-Aid for Scientific Research on Scientific Research B (No.\
16H03991 [ME and MMN] and No.\ 26287039 [MMN]), Scientific Research C
(No.\ 26400239 [TM]), Young Scientists B (No.\ 16K17681 [ME]) and
Innovative Areas (No.\ 16H06490 [TM] and 16H06492 [MMN]), and by World
Premier International Research Center Initiative (WPI Initiative),
MEXT, Japan.

\appendix

\section{Functional Determinant}
\label{app:funcdet}
\setcounter{equation}{0}

In this Appendix, we study Eq.\ \eqref{Gelfand-Yaglom}.  A
mathematical proof of a formula similar to Eq.\ \eqref{Gelfand-Yaglom}
has been provided in \cite{Kirsten:2003py, Kirsten:2004qv}.  Because
the set up adopted in \cite{Kirsten:2003py, Kirsten:2004qv} is
different from ours, we give a proof for the functional determinants
of ${\cal M}_J^{(S,L,\varphi)}$ and ${\cal M}_0^{(S,\varphi)}$.  For
the fluctuation operators of the transverse mode, Higgs mode, and FP
ghosts, similar argument holds and we can use Eq.\
\eqref{Gelfand-Yaglom}.  In this Appendix, ${\cal M}$ ($\widehat{\cal
  M}$) denotes ${\cal M}_J^{(S,L,\varphi)}$ or ${\cal
  M}_0^{(S,\varphi)}$ ($\widehat{\cal M}_J^{(S,L,\varphi)}$ or
$\widehat{\cal M}_0^{(S,\varphi)}$).

In order to explicitly impose the boundary condition to the
eigenfunctions, we first consider the functional determinant for the
eigenfunctions defined in the finite interval $0< r\leq R$; at the end
of the calculation, we take $R\rightarrow\infty$.  Let us denote the
$n$-th eigenvalues of ${\cal M}$ and $\widehat{\cal M}$ as $\lambda_n$
and $\widehat{\lambda}_n$, respectively.  Then, the corresponding
eigenfunctions, denoted as $\psi_n$ and $\widehat{\psi}_n$,
respectively, satisfy
\begin{align}
  {\cal M}\psi_n = &\, \lambda_n \psi_n,
  \\
  \widehat{\cal M} \widehat{\psi}_n = &\, 
  \widehat{\lambda}_n \widehat{\psi}_n.
\end{align}
with
\begin{align}
  \psi_n (R) = \widehat{\psi}_n (R) = 0.
  \label{BC:r=R}
\end{align}
Notice that we are interested in the functional determinant of
Hermitian operators, and hence $\lambda_n$ and $\widehat{\lambda}_n$
are real.

We start with introducing the $\zeta$ functions:
\begin{align}
  \zeta_{\cal M} (s) \equiv&\, \sum_n \lambda_n^{-s},
  \\
  \zeta_{\widehat{\cal M}} (s) \equiv&\, \sum_n \widehat{\lambda}_n^{-s}.
\end{align}
Then, the ratio of the functional determinants is defined as
\begin{align}
  \frac{\mbox{Det} {\cal M}}
  {\mbox{Det} \widehat{\cal M}}
  = e^{-[\zeta_{\cal M}' (0)-\zeta_{\widehat{\cal M}}' (0)]}.
  \label{detvszeta'(0)}
\end{align}
For the calculation of the $\zeta$ functions defined above, we
introduce the function $u(r;\lambda)$, which satisfies
\begin{align}
  {\cal M} u(r;\lambda) = \lambda u(r;\lambda),
  \label{M*u=lambda*u}
\end{align}
where $\lambda$ is a complex constant, and $u(r\rightarrow 0;\lambda)$
is required to be finite.

We first show that there are three (two) independent choices of
$u(r;\lambda)$ for $J\neq 0$ ($J=0$).  To see this, we expand
$u(r;\lambda)$ as
\begin{align}
  u(r;\lambda) = r^\nu \sum_{p=0}^{\infty} c_p r^p,
\end{align}
where $\nu$ is a non-negative constant because of the regularity
at the origin, and $c_p$ are constant 3-component (2-component)
objects for $J\neq 0$ ($J=0$) satisfying $c_0\neq 0$.  In order for
$u(r;\lambda)$ to obey Eq.\ \eqref{M*u=lambda*u}, the following
relation should hold:
\begin{align}
  {\cal M}^{\star} r^\nu c_0 = 0,
\end{align}
where ${\cal M}^{\star}$ is obtained from ${\cal M}$ by taking
$\bar{\phi}\rightarrow 0$, $\bar{\phi}'\rightarrow 0$, and
$\Delta_0\bar{\phi}\rightarrow 0$; it gives terms with the lowest
power in $r$ (i.e., terms of $O(r^{\nu-2})$) in Eq.\
\eqref{M*u=lambda*u}.  The above equation has solutions if
$\mbox{det}[r^{2-\nu}{\cal M}^{\star}r^\nu]=0$.  (Notice that
$r^{2-\nu}{\cal M}^{\star}r^\nu$ is a constant $3\times 3$ ($2\times
2$) matrix for $J\neq 0$ ($J=0$).)  It gives the following values of
$\nu$:
\begin{align}
  \nu = 
  \left\{
    \begin{array}{ll}
      2J-1,\, 2J,\, 2J+1 &: J\neq 0 \\
      0,\, 1 &: J= 0
    \end{array}
  \right. .
\end{align}
We can repeat the above argument for $\widehat{\cal M}$ to define
the function $\widehat{u}(r;\lambda)$, which satisfies 
\begin{align}
  \widehat{\cal M}\widehat{u}(r;\lambda) = 
  \lambda\widehat{u}(r;\lambda).
  \label{Mhat*u=lambdahat*u}
\end{align}
We can see that $u(r\rightarrow 0;\lambda)$ and
$\widehat{u}(r\rightarrow 0;\lambda)$ have the same power-law behavior
at $r\rightarrow 0$.  We choose the boundary conditions as follows:
for $J\neq 0$,
\begin{align}
  u_1 (r\rightarrow 0;\lambda) \simeq 
  \widehat{u}_1 (r\rightarrow 0;\lambda) \simeq
  &\,
  \left( \begin{array}{c}
      2J r^{2J-1} \\ L r^{2J-1} \\ 0
    \end{array}
  \right),
  \label{bc:u1}
  \\[2mm]
  u_2 (r\rightarrow 0;\lambda) \simeq 
  \widehat{u}_2 (r\rightarrow 0;\lambda) \simeq
  &\,
  \left( \begin{array}{c}
      0 \\ 0 \\ r^{2J}
    \end{array}
  \right),
  \label{bc:u2}
  \\[2mm]
  u_3 (r\rightarrow 0;\lambda) \simeq 
  \widehat{u}_3 (r\rightarrow 0;\lambda) \simeq
  &\,
  \left( \begin{array}{c}
      \displaystyle{ \frac{(J+1)\xi-J}{2L^2} r^{2J+1} }
      \\[3mm]
      \displaystyle{ \frac{(J+1)\xi-(J+2)}{4 L (J+1)} r^{2J+1} }
      \\[3mm]
      0
    \end{array}
  \right),
  \label{bc:u3}
\end{align}
and  for $J=0$, 
\begin{align}
  u_1 (r\rightarrow 0;\lambda) \simeq 
  \widehat{u}_1 (r\rightarrow 0;\lambda) \simeq
  &\,
  \left( \begin{array}{c}
      0 \\ 1
    \end{array}
  \right),
  \\[2mm]
  u_2 (r\rightarrow 0;\lambda) \simeq 
  \widehat{u}_2 (r\rightarrow 0;\lambda) \simeq
  &\,
  \left( \begin{array}{c}
      r \\ 0
    \end{array}
  \right).
\end{align}

\begin{figure}[t]
  \centerline{\epsfxsize=0.95\textwidth\epsfbox{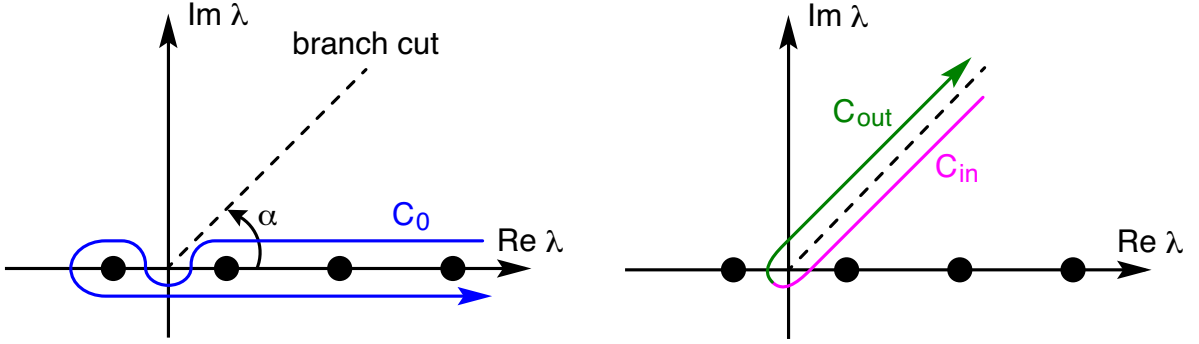}}
  \caption{\small The contours $C_0$, $C_{\rm in}$, and $C_{\rm out}$
    on the complex $\lambda$ plane.  The blobs on the real axis
    indicate the eigenvalues of the fluctuation operators.  The dotted
    line is the branch cut of $\lambda^{-s}$.}
  \label{fig:contour}
\end{figure}

Now we express the $\zeta$ functions using the functions $u_I$ and
$\widehat{u}_I$.  To make our argument explicit, we concentrate on the
case of $J\neq 0$; similar argument holds for the case of $J=0$.  For
the calculation the $\zeta$ functions, we use the following relation:
\begin{align}
  &
  \mbox{det} ( u_1 (r=R;\lambda=\lambda_n)~
  u_2 (r=R;\lambda=\lambda_n)~
  u_3 (r=R;\lambda=\lambda_n) )= 0,
  \\
  &
  \mbox{det} (\widehat{u}_1 (r=R;\lambda=\widehat{\lambda}_n)~
  \widehat{u}_2 (r=R;\lambda=\widehat{\lambda}_n)~
  \widehat{u}_3 (r=R;\lambda=\widehat{\lambda}_n) )= 0,
\end{align}
which are based on the boundary conditions on the eigenfunctions of
the differential operators (see Eq.\ \eqref{BC:r=R}).  Thus, the
logarithmic derivatives of above determinants with respect to
$\lambda$ have simple poles with unit residue at the eigenvalues of
corresponding fluctuation operators, and hence we can express the
$\zeta$ functions as \cite{Kirsten:2003py, Kirsten:2004qv}
\begin{align}
  \zeta_{\cal M} (s) - \zeta_{\widehat{\cal M}} (s) =
  \frac{1}{2\pi i}
  \int_{C_0} d \lambda \lambda^{-s}
  \frac{d}{d\lambda}
  \ln 
  \frac
  {\mbox{det} ( u_1 (R;\lambda)~
    u_2 (R;\lambda)~
    u_3 (R;\lambda) )}
  {\mbox{det} ( \widehat{u}_1 (R;\lambda)~
    \widehat{u}_2 (R;\lambda)~
    \widehat{u}_3 (R;\lambda) )},
  \label{zeta-intform}
\end{align}
where $C_0$ is a contour along the real axis, surrounding all
eigenvalues in counterclockwise direction (see Fig.\ \ref{fig:contour}).
The contour $C_0$ avoids the branch cut of $\lambda^{-s}$, which is
defined to be a straight line starting from the origin with the angle
$\alpha$ to the real axis.

The next task is to deform the contour to enclose the branch cut of
$\lambda^{-s}$.  To do so, information about the behavior of the
integrand at $|\lambda|\rightarrow\infty$ is necessary.  We can use
the fact that $u_I(r;\lambda)$ should satisfy
\begin{align}
  u_I(r;\lambda) = u_I^{\star}(r) +
  \int^r dr_1 r_1^3 G (r,r_1;\lambda) \delta {\cal M} (r_1) 
  u_I(r_1;\lambda),
  \label{IntegralEq} 
\end{align}
where $\delta {\cal M}\equiv{\cal M}-{\cal M}^{\star}$.  Here, the
function $G$ is given by
\begin{align}
  G (r_1,r_2;\lambda) \equiv &\,
  \frac{\pi \lambda \xi^2}{2}
  \left[
    v_1 (r_1;\lambda) w_1^{T} (r_2;\lambda)
    - w_1 (r_1;\lambda) v_1^{T} (r_2;\lambda)
  \right]
  \nonumber \\ &\, +
  \frac{\pi \lambda}{2}
  \left[
    v_2 (r_1;\lambda) w_2^{T} (r_2;\lambda)
    - w_2 (r_1;\lambda) v_2^{T} (r_2;\lambda)
  \right]
  \nonumber \\ &\, +
  \frac{\pi \lambda}{2}
  \left[
    v_3 (r_1;\lambda) w_3^{T} (r_2;\lambda)
    - w_3 (r_1;\lambda) v_3^{T} (r_2;\lambda)
  \right],
\end{align}
where
\begin{align}
  v_1 (r;\lambda) \equiv &\,
  \left( \begin{array}{c}
      \partial_r \\[2mm]
      \displaystyle{ \frac{L}{r} } \\[2mm]
      0
    \end{array}
  \right)
  \frac{J_{2J+1}(\sqrt{\lambda\xi}r)}{\lambda\xi r},
  \\[2mm]
  v_2 (r;\lambda) \equiv &\,
    \left( \begin{array}{c}
      \displaystyle{ \frac{L}{r} }
      \\[2mm]
      \displaystyle{ \frac{1}{r^2} \partial_r r^2 }
      \\[2mm]
      0
    \end{array}
  \right)
  \frac{J_{2J+1}(\sqrt{\lambda}r)}{\lambda r},
  \\[2mm]
  v_3 (r;\lambda) \equiv &\,
  \left( \begin{array}{c}
      0 \\ 0 \\ 1
    \end{array}
  \right)
  \frac{J_{2J+1}(\sqrt{\lambda}r)}{\sqrt{\lambda} r},
\end{align}
with $J_{2J+1}$ being the Bessel function of the first kind, and $w_I$
are obtained from $v_I$ by replacing $J_{2J+1}\rightarrow N_{2J+1}$
(with $N_{2J+1}$ being the Bessel function of the second kind).
Furthermore, $u^{\star}_I$ are solutions of the following differential
equation:
\begin{align}
  \left(
    {\cal M}^{\star} - \lambda
  \right) u^{\star}_I = 0,
\end{align}
which are regular at $r\rightarrow 0$.  Requiring the boundary
conditions given in Eq.\ \eqref{bc:u1}, \eqref{bc:u2}, and
\eqref{bc:u3} for $u^{\star}_1$, $u^{\star}_2$, and $u^{\star}_3$,
respectively,
\begin{align}
  u^{\star}_1 (r) = &\, 
  4 \Gamma (2J+2) \left( \frac{\sqrt{\xi \lambda}}{2} \right)^{-(2J-1)}
  v_1 (r;\lambda),
  \\[2mm]
  u^{\star}_2 (r) = &\,
  2 \Gamma (2J+2) \left( \frac{\sqrt{\lambda}}{2} \right)^{-2J}
  v_3 (r;\lambda),
  \\[2mm]
  u^{\star}_3 (r) = &\, 
  \frac{ 4 \Gamma (2J+2) }{\lambda L}
  \left( \frac{\sqrt{\lambda}}{2} \right)^{-(2J-1)}
  v_2 (r;\lambda)
  - 
  \frac{ 2 \Gamma (2J+2) }{\lambda J}
  \left( \frac{\sqrt{\xi \lambda}}{2} \right)^{-(2J-1)}
  v_1 (r;\lambda).
\end{align}

With large enough $\lambda$, we can see that the second term in the
right-hand side of Eq.\ \eqref{IntegralEq} is at most
$O(|\lambda|^{-1/2})$ compared to the first term.  This can be
understood by rewriting Eq.\ \eqref{IntegralEq} with a new
dimensionless variable $\rho\equiv\sqrt{|\lambda|}r$; with such a new
variable, $\delta{\cal M}$ can be treated as a perturbation and we can
expand the solution with respect to $\bar{\phi}_{\rm
  max}/\sqrt{|\lambda|}$ where $\bar{\phi}_{\rm max}$ is the scale of
the maximal amplitude of the bounce configuration.  Thus, we expect
that
\begin{align}
  u_I (r;\lambda\rightarrow\infty) \simeq &\,
  u^{\star}_I (r) + O(\lambda^{-1/2}).
  \label{u(rinf)}
\end{align}
A similar analysis applies to $\widehat{u}_I$, which results in
\begin{align}
  \widehat{u}_I (r;\lambda\rightarrow\infty) \simeq &\,
  u^{\star}_I (r) + O(\lambda^{-1/2}).
  \label{uhat(rinf)}
\end{align}
Thus, $u_I$ and $\widehat{u}_I$ have almost the same functional form
when $\lambda\rightarrow\infty$ and, for $\mbox{Im}\lambda\neq 0$,
\begin{align}
  \left.
    \frac{d}{d\lambda}
    \ln 
    \frac
    {\mbox{det} ( u_1 (R;\lambda)~
      u_2 (R;\lambda)~
      u_3 (R;\lambda) )}
    {\mbox{det} ( \widehat{u}_1 (R;\lambda)~
      \widehat{u}_2 (R;\lambda)~
      \widehat{u}_3 (R;\lambda) )}
  \right|_{\lambda\rightarrow\infty}
  \sim O(\lambda^{-3/2}).
  \label{lambda3/2}
\end{align}
This implies that, when $\mbox{Re}s>-\frac{1}{2}$, the integration at
$\lambda\rightarrow\infty$ does not contribute.\footnote
{For the calculation of $\zeta_{\cal M} (s) - \zeta_{\widehat{\cal M}}
  (s)$, convergence of the integration around $\lambda\sim 0$ requires
  $\mbox{Re}s<1$.}

Based on the above consideration, we can replace
$\int_{C_0}\rightarrow\int_{C_{\rm in}+C_{\rm out}}$, where the
contours $C_{\rm in}$ and $C_{\rm out}$ are those along the branch cut
incoming to and outgoing from the origin (see Fig.\
\ref{fig:contour}).  Using the fact that the functions $u_I$ and
$\widehat{u}_I$ are continuous at the branch cut of $\lambda^{-s}$,
\begin{align}
  \zeta_{\cal M} (s) - \zeta_{\widehat{\cal M}} (s) =
  e^{is(\pi-\alpha)}
  \frac{\sin (\pi s)}{\pi}
  \int_0^\infty d \lambda \lambda^{-s}
  \frac{d}{d\lambda}
  \ln 
  \frac
  {\mbox{det} ( u_1 (R;e^{i\alpha}\lambda)~
    u_2 (R;e^{i\alpha}\lambda)~
    u_3 (R;e^{i\alpha}\lambda) )}
  {\mbox{det} ( \widehat{u}_1 (R;e^{i\alpha}\lambda)~
    \widehat{u}_2 (R;e^{i\alpha}\lambda)~
    \widehat{u}_3 (R;e^{i\alpha}\lambda) )}.
\end{align}
Combining the above result with Eq.\ \eqref{detvszeta'(0)}, and taking
$R\rightarrow\infty$, we come to the most important formula in this
Appendix:\footnote
{Instead of Eq.\ \eqref{BC:r=R}, we may take an alternative boundary
  condition: $\psi'_n(R)=\widehat{\psi}'_n(R)= 0$.  A similar argument
  holds for this boundary condition; the result is given by Eq.\
  \eqref{DetM/DetMhat} with replacing $u_I\rightarrow u'_I$ and
  $\widehat{u}_I\rightarrow\widehat{u}'_I$.}
\begin{align}
  \frac{\mbox{Det} {\cal M}}{\mbox{Det} \widehat{\cal M}}
  = 
  \frac
  {\mbox{det} ( u_1 (r_\infty;\lambda=0)~
    u_2 (r_\infty;\lambda=0)~
    u_3 (r_\infty;\lambda=0) )}
  {\mbox{det} ( \widehat{u}_1 (r_\infty;\lambda=0)~
    \widehat{u}_2 (r_\infty;\lambda=0)~
    \widehat{u}_3 (r_\infty;\lambda=0) )}.
  \label{DetM/DetMhat}
\end{align}
Notice that, for the convergence of the above quantity, ${\cal M}$ and
$\widehat{\cal M}$ should have the same asymptotic behavior at
$r\rightarrow\infty$, which holds for the case of our interest.  In
the discussion given in Sections \ref{sec:nonzerovev} and
\ref{sec:zerovev}, the initial conditions of the solutions of Eqs.\
\eqref{M*u=lambda*u} and \eqref{Mhat*u=lambdahat*u} are taken to be
different.  (However, notice that we take three independent solutions
for ${\cal M}_J^{(S,L,\varphi)}$ and $\widehat{\cal
  M}_J^{(S,L,\varphi)}$, which are linear combinations of $u_I$ and
$\widehat{u}_I$, respectively.)  Then, we should use Eq.\
\eqref{Gelfand-Yaglom}.

\section{Solutions of Inhomogeneous Differential Equation}
\label{app:asymptotic}
\setcounter{equation}{0}

In this Appendix, we discuss the asymptotic behavior of the
inhomogeneous differential equation of the following form:
\begin{align}
  (\Delta_J - m^2) F(r) = S (r),
\end{align}
where $m$ is a constant, and the source term $S$ behaves as
\begin{align}
  S (r\rightarrow\infty) \simeq s_0 r^{-p} e^{\lambda r}.
  \label{sourceterm}
\end{align}
Using the modified Bessel functions, $I_{2J +1}$ and $K_{2J +1}$, the
solution of the above equation is given by
\begin{align}
  F(r) = &
  c_I \frac{I_{2J +1} (m r)}{r} 
  + c_K \frac{K_{2J +1} (m r)}{r}
  \nonumber \\ &
  + \frac{1}{r} 
  \left[ 
    I_{2J +1} (m r) \int^r dr' r'^2 K_{2J +1} (m r') S (r')
    - K_{2J +1} (m r) \int^r dr' r'^2 I_{2J +1} (m r') S (r')
  \right],
  \label{F(r)}
\end{align}
where $c_I$ and $c_K$ are constants.

Using the properties of the modified Bessel functions, i.e., $I_{2J
  +1}(z\rightarrow\infty)\simeq (2\pi z)^{-1/2}e^z$ and $K_{2J
  +1}(z\rightarrow\infty)\simeq (\pi/2 z)^{1/2}e^{-z}$, the asymptotic
behavior of the function $F(r)$ with the source term given in Eq.\
\eqref{sourceterm} can be expressed as\footnote
{Notice that, with the asymptotic behaviors of the modified Bessel
  functions, the integrations in Eq.\ \eqref{F(r)} can be expressed by
  using the incomplete gamma function as
  \begin{align*}
    \int^r_{r_0} dr' r'^2 I_{2J +1} (m r') S (r')  \simeq &\,
    \frac{1}{\sqrt{2\pi m}} (-m-\lambda)^{q}
    \left[ 
      \Gamma (q, -(m+\lambda)r_0)
      - \Gamma (q, -(m+\lambda)r)
    \right],
    \\
    \int^r_{r_0} dr' r'^2 K_{2J +1} (m r') S (r')  \simeq &\,
    \sqrt{\frac{\pi}{2m}} (m-\lambda)^{q}
    \left[ 
      \Gamma (q, (m+\lambda)r_0)
      - \Gamma (q, (m+\lambda)r)
    \right],
  \end{align*}
  with $q=(2p-5)/2$.  (Here, $m r_0\gg 1$ and $m
  r\gg 1$ are assumed.) In addition, in deriving Eq.\
  \eqref{F(r->inf)}, we also use $\Gamma(q,
  z\rightarrow\pm\infty)\simeq z^{q-1} e^{-z}$.  }
\begin{align}
  F (r\rightarrow\infty) \simeq
  c_I \frac{I_{2J +1} (m r)}{r}
  + c_K \frac{K_{2J +1} (m r)}{r}
  + \frac{s_0}{\lambda^2 - m^2} r^{-p} e^{\lambda r}.
  \label{F(r->inf)}
\end{align}
In the above expressions, the arbitrariness of the lower boundaries
of the integrations are absorbed into the constants $c_I$ and $c_K$.

\section{Functional Determinant with Small Perturbations}
\label{app:finiteorder}
\setcounter{equation}{0}

In this appendix, we outline the prescription to calculate $\delta
{\cal S}^{(S,L,\varphi)}_{\rm div}$ and $\delta {\cal S}^{(T)}_{\rm
  div}$ given in Eqs.\ \eqref{Sdiv_SLverphi} and \eqref{Sdiv_T},
respectively.  We expand the functional determinants with respect to
small perturbations, and calculate the functional determinant up to a
finite order of the perturbation.

The second term of $\delta {\cal S}^{(S,L,\varphi)}_{\rm div}$ as well
as $\delta {\cal S}^{(T)}_{\rm div}$ are described by the following
quantity with $N=2$:
\begin{align}
  \ell^{[N]} \equiv
  \left[ \ln
    \frac{\mbox{Det}(-\partial^2 + g^2 v^2 + g^2 \delta \bar{\phi}^2)}
    {\mbox{Det}(-\partial^2 + g^2 v^2)}
  \right]_{(g^2 \delta \bar{\phi}^2)^N},
  \label{L_PN}
\end{align}
where, as introduced in Section \ref{sec:renormalization},
$[\cdots]_{{\cal P}^N}$ indicates that the quantity in the square
bracket is evaluated up to $O({\cal P}^N)$.  Using the
angular-momentum decomposition, and also using Eq.\
\eqref{Gelfand-Yaglom}, the ratio of the functional determinants in
Eq.\ \eqref{L_PN} is given by
\begin{align}
  e^{\ell^{[N=\infty]}} = 
  \prod_J 
  \left[ 
    \frac{f_J(r_\infty)}{\widehat{f}_J(r_\infty)} 
  \right]^{(2J+1)^2},
\end{align}
where the functions $f_J$ and $\widehat{f}_J$ satisfy
\begin{align}
  &
  (\Delta_J - g^2 v^2 - g^2 \delta \bar{\phi}^2) f_J = 0,
  \label{eqforF}
  \\ &
  (\Delta_J - g^2 v^2) \widehat{f}_J = 0,
\end{align}
with $f_J(r\rightarrow 0)\simeq \widehat{f}_J(r\rightarrow 0)\simeq
r^{2J}$.  

Treating $\delta \bar{\phi}^2$ as a perturbation, we expand the
function $f_J$ as
\begin{align}
  f_J (r) = \sum_{n=0}^{\infty} f_J^{(n)} (r),
\end{align}
where $f_J^{(n)}$ is obtained by iteratively solving
\begin{align}
  (\Delta_J - g^2 v^2) f_J^{(n)} = g^2 \delta \bar{\phi}^2
  f_J^{(n-1)},
\end{align}
with $f_J^{(0)}=\widehat{f}_J$ and $f_J^{(n\neq 0)}(r\rightarrow
0)\sim O(r^{2J+1})$.

In order for the calculation of the counter terms for each angular
momentum, we decompose $\ell^{[N]}$ as
\begin{align}
  \ell^{[N]} = \sum_{J=0}^\infty \ell_J^{[N]}.
\end{align}
Then,
\begin{align}
  \ell_J^{[N=\infty]} = 
  (2J+1)^2
  \ln \left[
    \sum_{n=0}^{\infty} t^n
    \left( \frac{f_J^{(n)}(r_\infty)}{\widehat{f}_J(r_\infty)} \right)
  \right]_{t\rightarrow 1}.
\end{align}
For the calculation of $\ell_J^{[N]}$, we expand the right-hand side
of the above equation with respect to $t$, neglect terms of $O(t^p)$
with $p\geq N+1$, and take $t\rightarrow 1$.  In particular,
\begin{align}
  \ell^{[N=2]}_J =
  (2J+1)^2
  \left[ 
    \left( \frac{f_J^{(1)}(r_\infty)}{\widehat{f}_J(r_\infty)} \right)
    + \left( \frac{f_J^{(2)}(r_\infty)}{\widehat{f}_J(r_\infty)} \right)
    - \frac{1}{2} 
    \left( \frac{f_J^{(1)}(r_\infty)}{\widehat{f}_J(r_\infty)} \right)^2
  \right].
\end{align}

Eq.\ \eqref{Sdiv_SLverphi} contains functional determinants of the
fluctuation operators in the matrix form.  Even in such a case, we can
expand the functional determinant of our interest with respect to
perturbations.  For the calculation of the first term of the
right-hand side of Eq.\ \eqref{Sdiv_SLverphi}, we introduce $3\times
3$ functions ${\bf F}_J(r)$ for $J>0$, which are expanded as
\begin{align}
  {\bf F}_J (r) = \sum_{n=0}^{\infty}  {\bf F}_J^{(n)} (r).
\end{align}
They obey the following differential equations:
\begin{align}
  \left(
    \begin{array}{ccc}
      \displaystyle{ \Delta_J - \frac{3}{r^2} } - g^2 v^2
      & \displaystyle{ \frac{2L}{r^2} } 
      & 0 
      \\[3mm]
      \displaystyle{ \frac{2L}{r^2} } 
      & \displaystyle{ \Delta_J + \frac{1}{r^2} } - g^2 v^2
      & 0 
      \\[3mm]
      0
      & 0 
      & \Delta_J - \widehat{\Omega}
    \end{array}
  \right) {\bf F}_J^{(n)} 
  = 
  \left(
    \begin{array}{ccc}
      g^2 \delta\bar{\phi}^2 & 0 & 2 g \bar{\phi}'
      \\
      0 & g^2 \delta\bar{\phi}^2 & 0 
      \\
      2 g \bar{\phi}' & 0 & \delta\Omega 
    \end{array}
  \right) {\bf F}_J^{(n-1)},
  \label{diffeq_bfF}
\end{align}
for $n\geq 1$, and 
\begin{align}
  {\bf F}_J^{(0)} (r\rightarrow 0) \simeq
  \left(
    \begin{array}{ccc}
      2 J r^{2J-1}
      & -L r^{2J+1}
      & 0 
      \\
      L r^{2J-1}
      & 2 J r^{2J+1}
      & 0 
      \\
      0
      & 0 
      & r^{2J}
    \end{array}
  \right).
\end{align}
For $J=0$, ${\bf F}_{J=0}$ is a $2\times 2$ object; it satisfies
\begin{align}
  \left(
    \begin{array}{ccc}
      \displaystyle{ \Delta_0 - \frac{3}{r^2} } - g^2 v^2
      & 0 
      \\[3mm]
      0
      & \Delta_0 - \widehat{\Omega}
    \end{array}
  \right) {\bf F}_0^{(n)} 
  = 
  \left(
    \begin{array}{ccc}
      g^2 \delta\bar{\phi}^2 & 2 g \bar{\phi}'
      \\
      2 g \bar{\phi}' & \delta\Omega 
    \end{array}
  \right) {\bf F}_0^{(n-1)},
\end{align}
with 
\begin{align}
  {\bf F}_0^{(0)} (r\rightarrow 0) \simeq
  \left(
    \begin{array}{cc}
      r
      & 0 
      \\
      0 
      & 1
    \end{array}
  \right).
\end{align}
With the functions ${\bf F}_J$, we define
\begin{align}
  L_J^{[N=2]} \equiv &\, 
  (2J+1)^2 \mbox{tr} \left[ 
    \widehat{\bf F}_J^{-1} (r_\infty) {\bf F}_J^{(1)} (r_\infty)
    + \widehat{\bf F}_J^{-1} (r_\infty) {\bf F}_J^{(2)} (r_\infty)
  \right]
  \nonumber \\ &\, 
  - \frac{(2J+1)^2}{2} \mbox{tr} \left[
    \widehat{\bf F}_J^{-1} (r_\infty) {\bf F}_J^{(1)} (r_\infty)
    \widehat{\bf F}_J^{-1} (r_\infty) {\bf F}_J^{(1)} (r_\infty)
  \right],
\end{align}
where $\widehat{\bf F}_J\equiv{\bf F}_J^{(0)}$.  Then, the first term
of Eq.\ \eqref{Sdiv_SLverphi} is expressed as
\begin{align}
  \frac{1}{2} \left[
    \frac{\mbox{Det}(\widehat{\cal M}_{R_{\xi =1}}^{(A_\mu,\varphi)}
      + \delta {\cal M})}
    {\mbox{Det}\widehat{\cal M}_{R_{\xi =1}}^{(A_\mu,\varphi)}}
  \right]_{(\delta {\cal M})^2}
  =&\,
  \frac{1}{2} \sum_{J=0}^\infty L_J^{[N=2]}
  + \sum_{J=1/2}^\infty \ell_J^{[N=2]}.
\end{align}

In summary, the divergent parts given in Eqs.\ \eqref{Sdiv_SLverphi}
and \eqref{Sdiv_T} are given by
\begin{align}
  \delta {\cal S}^{(S,L,\varphi)}_{\rm div} = &\,
  \frac{1}{2} \sum_{J=0}^\infty L_J^{[N=2]}
  - \sum_{J=0}^\infty \ell_J^{[N=2]}
  - \ell_0^{[N=2]},
  \label{Sdiv_SLverphiWithL}
  \\
  \delta {\cal S}^{(T)}_{\rm div} = &\,
  \sum_{J=0}^\infty \ell_J^{[N=2]}.
  \label{Sdiv_TWithL}
\end{align}
Comparing the above equations with Eq.\ \eqref{Sdiv_SLverphi(J)} or
\eqref{Sdiv_T(J)}, we can obtain the counter terms for each angular
momentum, $s_J^{(S,L,\varphi)}$ and $s_J^{(T)}$.

\end{document}